\documentclass[amsmath,amssymb,rmp]{revtex4}
\AtBeginDvi{}
\usepackage{dcolumn}
\usepackage{bm}

\usepackage{epic}
\usepackage{eepic}
\usepackage{graphicx}
\usepackage{natbib}

\newcommand{\ket}[1]{|#1\rangle}

\renewcommand{\vec}[1]{{\bf{#1}}}

\newcommand{\Order}[1]{\mathcal{O}(#1)}

\begin{document}

\title{Quantum optics of ultra-cold molecules}

\author{D. Meiser}
\author{T. Miyakawa}
\author{H. Uys}
\author{P. Meystre}

\affiliation{Department of Physics, The University of Arizona,
    1118 E. 4th Street, Tucson, Arizona, 85705}

\begin{abstract}
Quantum optics has been a major driving force behind the rapid
experimental developments that have led from the first laser
cooling schemes to the Bose-Einstein condensation (BEC) of dilute
atomic and molecular gases. Not only has it provided
experimentalists with the necessary tools to create ultra-cold
atomic systems, but it has also provided theorists with a
formalism and framework to describe them: many effects now being
studied in quantum-degenerate atomic and molecular systems find a
very natural explanation in a quantum optics picture. This article
briefly reviews three such examples that find their direct
inspiration in the trailblazing work carried out over the years by
Herbert Walther, one of the true giants of that field.
Specifically, we use an analogy with the micromaser to analyze
ultra-cold molecules in a double-well potential; study the
formation and dissociation dynamics of molecules using the passage
time statistics familiar from superradiance and superfluorescence
studies; and show how molecules can be used to probe higher-order
correlations in ultra-cold atomic gases, in particular bunching
and antibunching.
\end{abstract}

\maketitle

\section{Introduction}

Quantum optics plays a central role in the physics of
quantum-degenerate atoms and molecules. Laser light and its
coherent and incoherent interactions with atoms are ubiquitous in
these experiments, and the tools that have culminated in the
achievement of Bose-Einstein condensation (BEC)
\cite{Cornell:BEC1995,Ketterle:BEC1995,Hulet:BEC1995} were first
studied and understood in quantum optics. Indeed, the deep
connection between quantum optics and cold atom physics was
realized well before the first experimental realizations of BEC,
both at the experimental and theoretical levels. On the theory
side, there are (at least) two important reasons why quantum
optics methods are well suited for the study of cold atoms
systems. First, bosonic fields have direct analogs in
electromagnetic fields, which have been extensively studied in
quantum optics. Second, for fermions the Pauli Exclusion Principle
restricts the occupation of a given mode to zero or one, and these
two states
--- mode occupied or empty --- can often be mapped onto a
two-level system, as we shall see. As a result, many situations
familiar from quantum optics are also found in cold-atom systems,
including matter-wave interference
\cite{Ketterle:BEC_interference}, atom lasers and matter-wave
amplifiers
\cite{Ketterle:BEC_coherence,Ketterle:Matter_wave_amplification,Kozuma:matter_wave_amplification,Law:matter_wave_amplification},
matter-wave beam splitters \cite{Burgbacher:beam_splitter_bosons},
four-wave mixing
\cite{Lenz:kapitza_dirac,Moore:qu_optics_bec,Rojo:talbot_oscillations,Taka:fermionic_fourwavemixing,Henning:fermionic_fourwavemixing,Search:BCS_scattering,Meiser:fermion_light_fwm},
and Dicke superradiance
\cite{Ketterle:superradiance,Moore:superradiance}, to name a few.

At the same time the physics of ultra-cold atoms is much richer
than its quantum-optical counterpart since atoms can be either
fermions
\cite{DeMarco:degeneratefermions,DeMarco:degeneratefermions2,Hadzibabic:ultracold_fermions}
or bosons and have a rich internal structure. In addition, the
interaction between atoms can be tuned relatively easily on fast
time scales using for instance Feshbach resonances
\cite{Timmermans:Feshbachresonances,Stoof:FRreview,Inouye:MoleculeFR,Duerr:MoleculeRP,Inouye:heteromolecules,Stan:heteromolecules}
or two-photon Raman transitions
\cite{Wynar:MoleculePA,Theis:opticalFR}. Indeed, some of the most
exciting recent developments in the physics of ultra-cold atoms are
related to the coherent coupling of atoms to ultra-cold molecules
by means of Feshbach resonances
\cite{Regal:UltracoldMolecule,Duerr:MoleculeRP}, and
photo-association \cite{Wynar:MoleculePA,Kerman:polarMolecule}.
Both bosons and fermions have been successfully converted into
molecules. In both cases BEC of molecules has been observed
\cite{Donley:MoleculeBEC,Greiner:MoleculeBEC,Zwierlein:MolecularBEC,Jochim:MolecularBEC},
and the long-standing question of the BEC-BCS crossover is being
investigated experimentally and theoretically in those systems
\cite{Regal:BEC_BCS_crossover,Bartenstein:,Zwierlein:BEC_BCScrossover,Timmermans:BCS,Holland:resonant_superfluidity,Ohashi:BCSBEC}.
Other developments with close connections with quantum optics
include the trapping of atoms in optical lattices
\cite{Jaksch:BECinLattice,Bloch:MottInsulator1,Bloch:MottInsulator2},
which play a role closely related to a high-Q resonator in cavity
QED \cite{Walther:QEDexperiments,Search:micromaserarray} and leads
in addition to fascinating connections with condensed matter
physics and quantum information science.

With so many close connections between the physics of
quantum-degenerate atomic and molecular systems and quantum
optics, it is natural and wise to go back to the masters of that
field to find inspiration and guidance, and this is why Herbert
Walther's intellectual imprint remains so important. This brief
review illustrates this point with three examples. Section
\ref{micromaser} shows that the conversion of pairs of fermions
into molecules in a double-well potential can be described by a
generalized Jaynes-Cummings model. Using this equivalence, we show
that the dynamics of the molecular field at each site can be
mapped to that of a micromaser, one of Herbert Walther's most
remarkable contributions \cite{Meschede:oneatommaser}. Section
\ref{passagetimestatistics} further expands on the mapping of
ultra-cold fermion pairs onto two-level atoms to study the role of
fluctuations in the association and dissociation rates of
ultra-cold molecules. We show that this system is closely related
to Dicke superradiance, and with this analogy as a guide, we
discuss how the passage time fluctuations depend sensitively on
the initial state of the system. In a third example, inspired by
Herbert Walther's work on photon statistics and antibunching
\cite{Krause:numberstate,Rempe:subpoissonian1,Rempe:subpoissonian2,Brattke:numberstates}
section \ref{countingstatistics} analyzes how the statistics of
their constituent atoms affects the counting statistics of
molecules formed by photo-association. We compare the three cases
where the molecules are formed from a BEC, an ultra-cold Fermi gas
and a Fermi system with a superfluid component. The concept of
quantum coherence developed by R. J. Glauber and exploited in many
situations by H. Walther and his coworkers, in particular in their
studies of resonance fluorescence, are now applied to
characterizing the statistical properties of the coupled
atom-molecule system. Finally, section
\ref{diagnosingwithmolecules} further elaborates on these ideas to
probe spatial correlations and coherent properties of atomic
samples, and we find that the momentum distribution of the
molecules contains detailed information about the second-order
correlations of the initial atomic gas.

\section{\label{micromaser} Molecular micromaser}

Ultra-cold atoms and molecules trapped in optical lattices provide
an exciting new tool to study a variety of physics problems. In
particular, they provide remarkable connections with the condensed
matter of strongly correlated systems and with quantum information
science, a very well controlled environment to study processes
such as photo-association \cite{Ryu:AMRabioscillation}, and, from a
point-of-view more directly related to quantum optics, can be
thought of as matter-wave analog of photons trapped in high-$Q$
cavities. In particular, the high degree of real-time control of
the system parameters offers the opportunity to directly
experimentally study some of the long-standing questions of
condensed matter physics, such as the ground state structure of
certain models and many-body dynamic properties
\cite{Zoller:OLreview}. The coherent formation of molecules in an
optical lattice via either Feshbach resonances and two-photon
Raman photo-association has been studied both theoretically
\cite{Jaksch:creationmoleculeinOL,Damski:dipolarSFinOL,molmer:JaynesCummingsinOL,Moore:twospeciesMIinOL,Esslinger:atommoleculeinOL}
and experimentally
\cite{Rom:stateselectivemoleculeinOL,Kohl:fermionsin3DOL,Ryu:AMRabioscillation,Stoferle:Mol_OL_FR}.
In particular, the experiment of Ref. \cite{Ryu:AMRabioscillation}
observed reversible and coherent Rabi oscillations in a gas of
coupled atoms and molecules.

The idea of the molecular micromaser
\cite{Search:molecularmicromaser} relies on the observation that,
as a consequence of Fermi statistics, the photo-association of
fermionic atoms into bosonic molecules can be mapped onto a
generalized Jaynes-Cummings model. This analogy allows one to
immediately translate many of the results that have been obtained
for the Jaynes-Cummings model to atom-molecule systems. In
addition, the molecular system possesses several properties that
have no counterpart in the quantum optics analog, giving rise to
interesting generalizations of the original micromaser problem
\cite{Meschede:oneatommaser,Filipowicz:Theorymicromaser,Guzman:semiclassicalmicromaser,Rempe:photonstatistics}.
One of these new features is the inter-site tunneling of atoms and
molecules between adjacent lattice sites, leading to a system that
can be thought of as an array of molecular micromasers
\cite{Search:micromaserarray}.

To see how this works, rather than treating a full lattice
potential we consider the dynamics of the molecular field in the
simpler model of a coupled atom-molecule system in a double-well
potential. We first show that inter-well tunneling enhances number
fluctuations and eliminates trapping states in a manner similar to
thermal fluctuations.  We also examine the buildup of the relative
phase between the two molecular states localized at the two wells
due to the combined effect of inter-well tunneling and two-body
collisions. We identify three regimes, characterized by different
orders of magnitude of the ratio of the two-body collision
strength to the inter-well tunneling coupling. The crossover of
the non-equilibrium steady state from a phase-coherent regime to a
phase-incoherent regime is closely related to the phase locking of
condensates in Josephson-type
configurations~\cite{Leggett:ReviewBEC}, while we consider an open
quantum system with incoherent pump and molecular loss which
results in a dissipative steady state.

\subsection{Model}

We consider a mixture of two hyperfine spin states
$|\sigma=\uparrow,\downarrow\rangle$ of fermionic atoms of mass
$m_f$ trapped in a double-well potential at temperature $T=0$,
which can be coherently combined into bosonic molecules of mass
$m_b$ via two-photon Raman photo-association. If the band-gap of
the lattice potential is much larger than any other energy scale
in the system, the fermions and molecules occupy only the lowest
energy level of each well and the number of fermions of a given
spin state is at most one in each well.

In the tight binding approximation, the effective Hamiltonian
describing the coupled atom-molecule system is
\begin{equation}
\hat{H}=\sum_{i=l,r} (
\hat{H}_{0i}+\hat{H}_{Ii})+\hat{H}_T,
\end{equation}
where
\begin{eqnarray}
\hat{H}_{0i} &=&
\hbar(\omega_b+\delta)\hat{n}_{i}+\hbar\omega_f
(\hat{n}_{\uparrow i}+\hat{n}_{\downarrow i})+\frac{1}{2}\hbar U_{b}\hat{n}_{i}(\hat{n}_{i}-1)
\nonumber\\
&+&\hbar U_{x}\hat{n}_{i}(\hat{n}_{\uparrow i}+\hat{n}_{\downarrow i})+\hbar
U_{f}\hat{n}_{\uparrow i}\hat{n}_{\downarrow i},\\
\hat{H}_{Ii} &=&\hbar\chi(t)
\hat{b}^{\dagger}_i\hat{c}_{\uparrow i}\hat{c}_{\downarrow i}+H.c.,\\
\hat{H}_T &=&
-\hbar J_b \hat{b}^{\dagger}_l\hat{b}_r -\hbar
J_f(\hat{c}^{\dagger}_{\uparrow l}\hat{c}_{\uparrow r}
+\hat{c}^{\dagger}_{\downarrow l}\hat{c}_{\downarrow r})+H.c.
\end{eqnarray}
Here $\hat{c}_{\sigma i}$ and $\hat{b}_i$, $i=l,r$, are the
annihilation operators of fermionic atoms and bosonic molecules in
the left ({\it l}) and right ({\it r}) wells, respectively. The
corresponding number operators $\hat{n}_{i}=
\hat{b}^{\dagger}_i\hat{b}_i$ and $\hat{n}_{\sigma
i}=\hat{c}^{\dagger}_{\sigma i}\hat{c}_{\sigma i}$ have
eigenvalues $n_{i}$ and $n_{\sigma i}$, respectively, and
$\hbar\omega_b$ and $\hbar\omega_f$ are the energies of the molecules and
atoms in the isolated wells.

The terms proportional to $U_b$, $U_x$, and $U_f$ in
$\hat{H}_{0i}$ describe on-site two-body interactions between
molecules, between atoms and molecules, and between atoms,
respectively. The interaction Hamiltonian $\hat{H}_{Ii}$ describes
the conversion of atoms into molecules via two-photon Raman
photo-association. The photo-association coupling constant $\chi(t)$
is proportional to the far off-resonant two-photon Rabi frequency
associated with two nearly co-propagating
lasers~\cite{Heinzen:superchemistry}, and $\delta$ is the
two-photon detuning between the lasers and the energy difference
of the atom pairs and the molecules. The tunneling between two
wells is described by the parameters $J_b$ and $J_f$ in the
tunneling Hamiltonian $\hat{H}_T$.

The molecular field is ``pumped'' by a train of short
photo-association pulses of duration $\tau$, separated by long
intervals $T \gg \tau$ during which the molecules are subject only
to two-body collisions and quantum tunneling between the
potential wells, as well as to losses due mainly to three-body
collisions and collisional relaxation to low-lying vibrational
states. In the absence of inter-well tunneling, this separation
of time scales leads to a situation very similar to that
encountered in the description of traditional micromasers, with
the transit of individual two-level atoms through the micromaser
cavity replaced by the train of photo-association pulses.

The dynamics of the molecular field in the double-well system is
governed by the following four mechanisms:

(i) Coherent pumping by injection of pairs of fermionic atoms
inside the double-well.  This process is the analog of the
injection of two-level atoms into a micromaser cavity.  The
injection of pairs of fermionic atoms into the double-well
potential can be accomplished e.g. by Raman transfer of atoms from
an untrapped internal state \cite{Mandel:coherenttransport,
Jaksch:BECinLattice}. This results in the pumping of fermions into
the double well at a rate $\Gamma$~\cite{Search:inputoutput}. We
assume that for times $T \gg \Gamma^{-1}$, a pair of fermions has
been transferred to the two wells with unit probability, that is,
the state of the trapped fermions in well $i$ is
\begin{equation}
 |e_i\rangle= \hat{c}^\dagger_{\downarrow i}\hat{c}^\dagger_{\uparrow i}|0\rangle.
\end{equation}

(ii) Molecular damping, which is the analog of cavity damping.
During the time intervals T when the photo-association lasers are
off, the molecular field decays at rate $\gamma$
\cite{Search:molecularmicromaser}. The decay of the molecules is
due to Rayleigh scattering from the intermediate molecular excited
state, three-body inelastic collisions between a molecule and two
fermions, and collisional relaxation from a vibrationally excited
molecular state to deeply bound states. These loss mechanisms can
be modeled by a master equation, see e.g.
\cite{miyakawa:doublewellmicromasers,Meystre:elements_of_quantum_optics,Scully:Quantum_Optics}.

(iii) The application of a train of photo-association pulses.
This mechanism is formally analogous to the Jaynes-Cummings interaction
between the single-mode field and a sequence
of two-level atoms traveling through the microwave cavity in the
conventional micromaser. As already mentioned, we assume that
these are square pulses of duration $\tau$ and period $T+\tau$,
with $\tau$ much shorter than all other time scales in this model,
$\tau\ll J_{b,f}^{-1},\gamma^{-1}$. This assumption is essential
if we are to neglect damping and tunneling while the
photo-association fields are on. The change in the molecular field
resulting from atom-molecule conversion is given by
\begin{equation}
F_i(\tau)\hat{\rho}_b\equiv
Tr_{a}[U_i(\tau)\hat{\rho}_{ab}(t) U^\dagger_i(\tau)],
\end{equation}
where $\hat{\rho}_{ab}$ is the total density operator of the
atom-molecule system and $Tr_a[$ $]$ denotes the trace over the
atomic variables, $U_i(\tau)=\exp{(-i\hat{h}_i\tau/\hbar)}$ being
the evolution operator for a single-well Hamiltonian,
$$
    \hat{h}_{i}=\hat{H}_{0i}+\hat{H}_{Ii}.
$$
The key observation that allows us here and below to build a bridge
from the cold atoms and molecular system to quantum optics systems
is that by means of the mapping \cite{Anderson:RPAinSuperconductivity}
\begin{eqnarray}
\hat{\sigma}_{-i}&=&c_{\uparrow i}c_{\downarrow i},\nonumber \\
\hat{\sigma}_{+i}&=&c^\dagger_{\downarrow i}c^\dagger_{\uparrow i},\nonumber \\
\hat{\sigma}_{zi}&=&c^\dagger_{\uparrow i} c_{\uparrow i}+c^\dagger_{\downarrow i} c_{\downarrow i}-1,
    \label{andersonmapping}
\end{eqnarray}
the atomic degrees of freedom take the form of a fictitious two
level system. The operators
$\hat{\sigma}_{+i}$,$\hat{\sigma}_{-i}$, and $\hat{\sigma}_{zi}$
can be interpreted as the raising and lowering operators of the
fictitious two-level atom and the population difference,
respectively. The single-well Hamiltonian takes the form
\begin{eqnarray}
    \label{JCH}
    \hat{h}_i&=&\hbar\left(\omega_b+U_{x}\right)\hat{n}_{i}+
    \hbar\left(\omega_f+
    U_{x}\hat{n}_{i}\right)\hat{\sigma}_{zi}\hspace{2cm}\nonumber\\
    &+&\hbar\left(\chi(t)\hat{b}_i^{\dagger}\hat{\sigma}_{-i}+
    \chi^*(t)\hat{b}_i\hat{\sigma}_{+i} \right) +\frac{\hbar}{2}U_{b}\hat{n}_{i}(\hat{n}_{i}-1)
    \end{eqnarray}
where we have dropped constant terms and we have redefined
$\omega_b$ and $\omega_f$ according to $\omega_b+\delta\rightarrow
\omega_b$ and $\omega_f+U_{f}/2\rightarrow \omega_f$.

The Hamiltonian ${\hat h}_i$ is Jaynes-Cummings-like, and for
$\chi=\rm{const.}$, the resulting dynamics can be determined
within the two-state manifolds of each well $\{ |e_i,n_{i}\rangle
, |g_i,n_{i}+1\rangle \}$ by a simple extension of the familiar
solution to the Jaynes-Cummings model. Within each manifold the
system undergoes Rabi oscillations.

Since tunneling is neglected during the photo-association steps,
the two wells are independent of each other and identical to each
other. The resultant molecular gain is then modeled by
independent coarse-grained equations of motion for the reduced
density matrices of each molecular mode.

(iv) The unitary time evolution of the molecular field under the
influence of two-body collisions and quantum tunneling, a process
absent in conventional micromasers. During the intervals $T$ it is
governed by
\begin{equation}
\label{MolH}
\frac{\partial \hat{\rho}_b}{\partial t}=-\frac{i}{\hbar}[\hat{H}_b,\hat{\rho}_b],
\end{equation}
where the Hamiltonian
\begin{equation}
\label{reducedH}
\hat{H}_b=-\hbar J_b (\hat{b}_l^\dagger \hat{b}_r
+\hat{b}_r^\dagger \hat{b}_l)
+\hbar\frac{U_b}{4}(\hat{n}_{l}-\hat{n}_{r})^2
\end{equation}
contains tunneling and collisions.
In Eq.~(\ref{reducedH}), we have neglected terms that are
functions only of $\hat{N}=\hat{n}_l+\hat{n}_r$, a step justified
as long as the initial density matrix is diagonal in the total
number of molecules in the two wells.

Combining the coherent and incoherent processes (i) to (iv), we
obtain the full evolution of the molecular field
\begin{equation}
\label{cgmastereq}
\frac{\partial\hat{\rho}_b}{\partial t}=\sum_{l,r}{\mathcal L}_i\hat{\rho}_b
+\frac{1}{T}\sum_{l,r}\left[F_i(\tau)-I_i\right]\hat{\rho}_b
-\frac{i}{\hbar}\left[H_b,\hat{\rho}_b\right],
\end{equation}
where $\hat{\rho}_b$ is the reduced density matrix of the
molecules. The initial condition for the molecules is taken to be
the vacuum state. Because the molecular pumping and decay is the
same in both wells, the density matrix $\rho$ remains diagonal in
the total number of molecules in the two wells for all times.
This is a generalization of the micromaser result that the photon
density matrix will remain diagonal if it is initially diagonal in
a number state basis \cite{Filipowicz:Theorymicromaser}.

\subsection{Results}

The master equation describing the molecular micromaser dynamics
contains six independent parameters: the number of
photo-association cycles per lifetime of the molecule,
$N_{ex}=1/\gamma T$; the ``pump parameter''
$\Theta=\sqrt{N_{ex}}|\chi|\tau$; the two-body collision strength
and tunneling coupling strength per decay rate, $u_b=U_b/\gamma$
and $t_J=J_b/\gamma$; and finally, the detuning parameter
$\eta= (2\omega_f-\omega_b)/2|\chi|$
and the nonlinear detuning parameter
$\beta\equiv (2U_x-U_b)/2|\chi|$.

In our model, the atomic and molecular level separations in the wells
are required to be much larger than the relevant interaction energies,
\begin{equation}
\label{condition1} \hbar \omega_{b}\gg U_b \langle
\hat{n}_i\rangle(\langle\hat{n_{i}}\rangle-1),\,\,\,
|\chi|\sqrt{\langle\hat{n}_i\rangle},
\end{equation}
$\langle \hat{n}_i \rangle$ being the average number of molecules
in well $i$. A comparison with actual experimental parameters
\cite{Jaksch:BECinLattice,Bloch:MottInsulator1,miyakawa:doublewellmicromasers}
shows that these conditions are satisfied as long as the number of molecules
doesn't exceed 10. In addition, the neglect of inter-well tunneling and damping
effects during the photo-association pulses requires that
\begin{equation}
\label{condition2}
    \tau \ll J_b^{-1}, \gamma^{-1}.
\end{equation}
This condition is satisfied in typical experiments.

In the remainder of this section, we discuss the dynamics of the molecular
field obtained by direct numerical integration of the master equation with a
Runge-Kutta algorithm until a dissipative steady state is reached.  For
simplicity, we confine our discussion to exact resonance only, $\eta=\beta=0$,
and a fixed value of $N_{ex}=10$.

\subsubsection{single-well molecular statistics}

We first discuss the statistics of a single-well molecular mode,
which is given by tracing over the full density matrix with
respect to degrees of freedom of the other localized mode as
\begin{equation}
P(n_{l\{r\}})=Tr_{r\{l\}}[\rho(n_l,n_r;m_l,m_r)]=\sum_{n_{r\{l\}}}\rho(n_l,n_r;n_l,n_r).
\end{equation}
We note that off-diagonal elements of the density matrix for a
single well are zero. Since the initial state of the molecules in
each well is the same, i.e. the vacuum state, and $\hat{H}_b$ is
invariant with respect to the interchange $l \leftrightarrow r$,
the molecule statistics for left and right wells are identical,
$P(n_l)=P(n_r)\equiv P(n)$.

\begin{figure} \includegraphics[width=0.9\columnwidth]{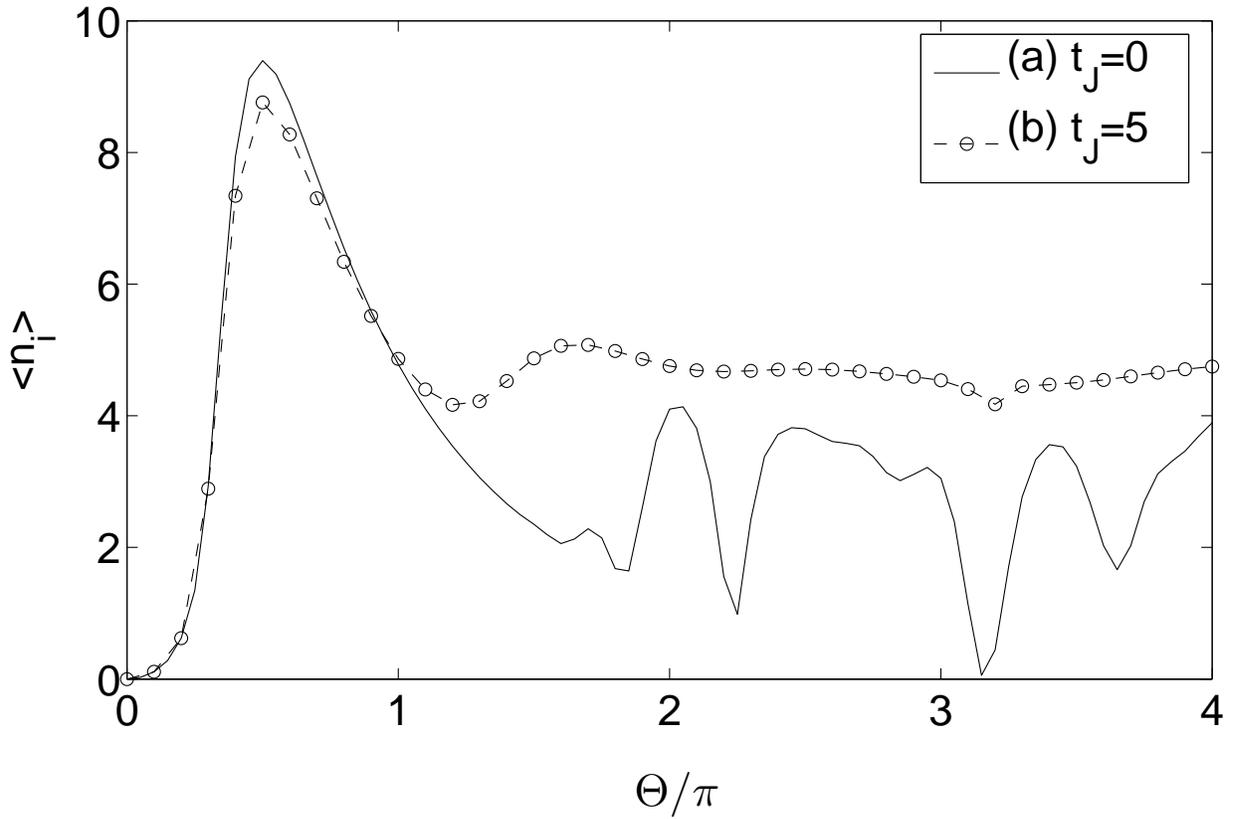}
       \caption{ $\langle \hat{n}_i \rangle$ versus $\Theta/\pi$ for $u_b=0$ and $N_{ex}=10$,
       and for (a) $t_J=0$ and (b) $t_J=5$.} \label{figII1}
     \end{figure}

\begin{figure} \includegraphics[width=0.9\columnwidth]{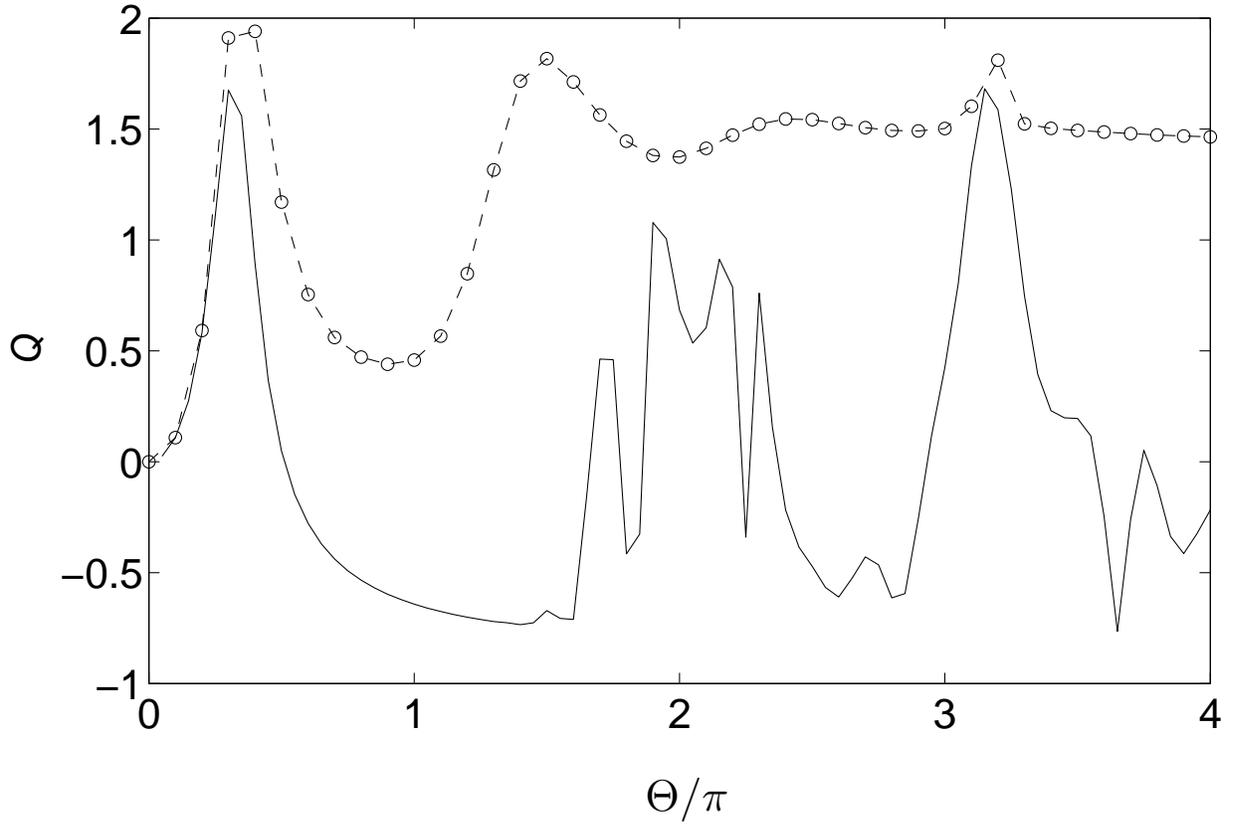}
       \caption{ {\it Q} parameter versus $\Theta/\pi$ for $u_b=0$ and $N_{ex}=10$,
     and for (a) $t_J=0$ and (b) $t_J=5$.} \label{figII2}
     \end{figure}

Figure~\ref{figII1} shows the steady-state average number $\langle
\hat{n}_i \rangle$ is plotted as a function of the pump parameter
$\Theta$ for $u_b=0$, $N_{ex}=10$. In the absence of inter-well
tunneling, corresponding to Fig.~\ref{figII1}(a), the result
reproduces that of conventional micromasers, with a ``lasing''
threshold behavior at around $\Theta\approx 1$ and an abrupt jump
to a higher mean occupation at about the first transition point,
$\Theta\simeq 2\pi$. The former effect is not affected by the
tunneling coupling as shown in Fig.~\ref{figII1}(b). However, the
latter abrupt jump disappears in the presence of interwell
tunneling. This is because the coupling to the other well leads to
fluctuations in the number of molecules in each well and has an
effect similar to thermal fluctuations in the traditional
micromaser theory. The enhancement of fluctuations can also be
seen in Fig.~\ref{figII2} where the Mandel ${\it Q}$-parameter
\[
{\it Q}=\frac{\langle \hat{n}_i^2 \rangle-\langle \hat{n}_i \rangle^2}{\langle \hat{n}_i \rangle}-1
\]
is plotted as a function of $\Theta$.

\begin{figure} \includegraphics[width=0.9\columnwidth]{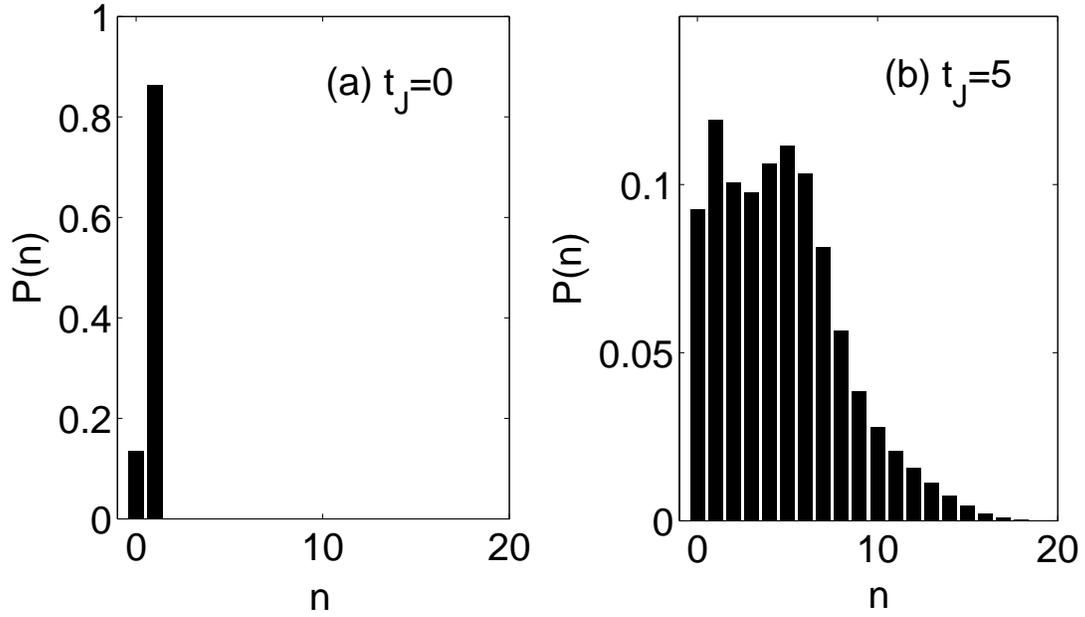}
       \caption{ Molecular number statistics $P(n_i)$ for $\Theta=\sqrt{5}\pi$,
$u_b=0$ and $N_{ex}$, and for (a) $t_J=0$ and (b) $t_J=5$.} \label{figII3}
     \end{figure}

It is known that in the usual micromaser the sharp resonance-like
dips in $\langle \hat{n}_i \rangle$ and ${\it Q}$ are attributable
to trapping states, which are characterized by a sharp photon
number. For the specific value of $\Theta=\sqrt{5}\pi$, as shown
in Fig.~\ref{figII3}(a), the number probability does not reach
beyond number state $|n_i=1\rangle$ in the case of $t_J=0$. As
shown in Fig.~\ref{figII3}(b), the tunneling coupling makes
possible transitions into higher number states and eliminates the
trapping state in a manner similar to thermal fluctuations in the
conventional micro maser.

\subsubsection{phase coherence between two micromasers with tunneling coupling}

So far we have discussed the single-well molecule statistics and
how it is affected by inter-well tunneling. Now we turn to a more
detailed discussion of the phase coherence between the two
localized modes. It is very useful to divide the parameter space
of the ratio of the two-body collision strength to the inter-well
tunneling coupling into three regimes \cite{Leggett:ReviewBEC}:
``Rabi-regime'' ($u_b/t_J \ll \langle \hat{N}\rangle^{-1}$);
``Josephson-regime'' ($\langle \hat{N} \rangle^{-1} \ll u_b/t_J
\ll\langle \hat{N} \rangle$); and ``Fock-regime'' ($\langle
\hat{N} \rangle \ll u_b/t_J$); where $\langle \hat{N} \rangle$
denotes the average total molecule number.

The analysis of the relative coherence of the molecular fields in
the two wells is most conveniently discussed in terms of the
angular momentum representation
\begin{eqnarray}
\hat{J}_{+}&=&\hat{J}_x+i\hat{J}_y=\hat{b}^\dagger_l \hat{b}_r,\nonumber \\
\hat{J}_{-}&=&\hat{J}_x-i\hat{J}_y=\hat{b}^\dagger_r \hat{b}_l,\nonumber \\
\hat{J}_z&=&\frac{1}{2}(\hat{b}_l^\dagger\hat{b}_l -\hat{b}_r^\dagger \hat{b}_r),
\nonumber \\
\hat{J}^2&=&\frac{\hat{N}}{2}\left(\frac{\hat{N}}{2}+1\right).
\end{eqnarray}
The symmetry of the density matrix with respect to the two wells
furthermore implies that $\langle \hat{J}_z\rangle
=\langle\hat{J}_y\rangle=0$. The first-order coherence between the
molecular fields in the left and right potential wells is then
given by $\langle {\hat J}_x\rangle$.
      \begin{figure} \includegraphics[width=0.9\columnwidth]{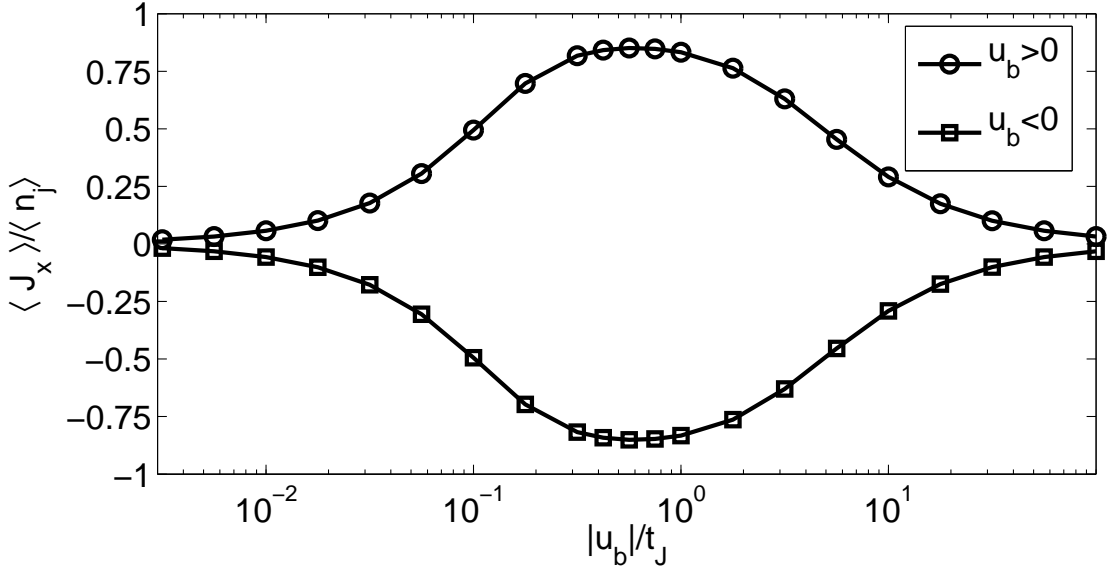}
    \caption{ $\langle \hat{J}_x \rangle/\langle n_j\rangle$ versus
      $u_b/t_J$ for $\Theta=\pi$ and $t_J=2.5$.} \label{figII4}
      \end{figure}
Figure~\ref{figII4} shows the normalized steady-state first-order
coherence $\langle\hat{J}_x\rangle/\langle \hat{n}_{j} \rangle$ as
a function of $u_b/t_J$ for $\Theta=\pi$ and $t_J=2.5$. $\langle
\hat{J}_x\rangle$ is suppressed in both the Rabi and Fock regimes
and has an extremum at $|u_b|/t_J\sim 0.6$. In the Fock regime,
$|u_b|/t_J\gg \langle \hat{N}\rangle \sim 10$, the nonlinearity in
$\hat{H}_b$ dominates and reduces the coherence between the
localized states of each well. We note that the average occupation
numbers for each well are relatively unaffected by $u_b/t_J$, with
$\langle\hat{n}_j\rangle=\langle\hat{N}\rangle/2=4.78-4.87$ for
$|u_b|/t_J=10^{2}-10^{-2.5}$.



The reason why the first-order coherence is suppressed in the weak coupling limit,
$u_b=0$, can be understood as follows.
The expectation value $\langle\hat{J}_x\rangle$ corresponds
to the difference in occupation numbers between the in-phase,
$\hat{b}_s=(\hat{b}_l+\hat{b}_r)/\sqrt{2}$, and out-of-phase,
$\hat{b}_a=(\hat{b}_l-\hat{b}_r)/\sqrt{2}$, states of the
localized states of each well, $\hat{J}_x=\hat{b}_s^\dagger
\hat{b}_s-\hat{b}_a^\dagger\hat{b}_a$.
Since the bandwidth of the photo-association pulse is
larger than their energy splitting, $1/\tau\gg J_b$,
those states are equally populated, resulting in
$\langle \hat{J}_x\rangle=0$ for $u_b=0$.
Thus, the origin of the mutual coherence between two molecular modes
is due solely to two-body collisions. Furthermore, we remark that
a semiclassical treatment results in $\langle \hat{J}_x\rangle=0$
for all times and all values of $u_b/t_J$
\cite{miyakawa:doublewellmicromasers}.
Hence, we conclude that the build-up of $\langle\hat{J}_x\rangle$
is a purely quantum-mechanical effect due to quantum fluctuations.

The phase distribution of the two wells can be studied using the Pegg-Barnett
phase states \cite{Pegg:unitaryphaseoperator,Barnett:opticalphasecorrelations,Javanainen:atomnumberfluctuations,Luis:phasedifferenceoperator}.
Since the density matrix is diagonal in the total number of molecules it is
sufficient to consider the relative phase.
      \begin{figure} \includegraphics[width=0.9\columnwidth]{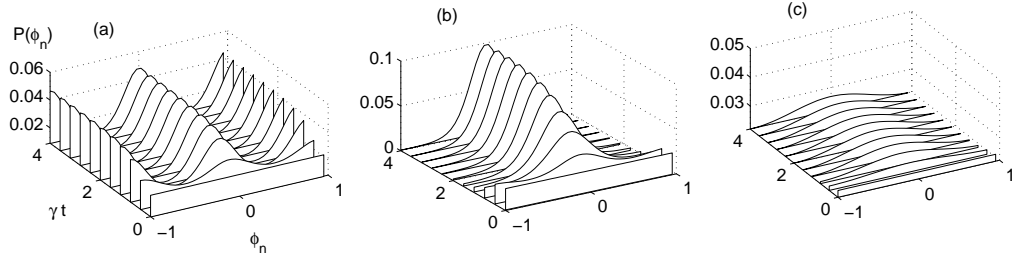}
      \caption{Time evolution of $P(\phi_n)$ for $\Theta=\pi$, $t_J=2.5$ and
      for (a) $u_b/t_J=0.0032$, (b) $u_b/t_J=0.5623$, (c) $u_b/t_J=56.23$.}
      \label{figII5}
    \end{figure}
Figure~\ref{figII5} shows the time evolution of the relative phase
distribution in three different regimes:
(a) Rabi, $u_b/t_J=0.0032$,
(b) Josephson, $u_b/t_J=0.5623$, and
(c) Fock $u_b/t_J=56.23$,
for $\Theta=\pi$, $t_J=2.5$.
Since the vacuum state is taken as the initial state, the relative
phase at $t=0$ is randomly distributed, $P(\phi_n)=1/(s+1)$.

In the Rabi regime, corresponding to Fig.\ref{figII5}(a), bimodal
phase distribution with peaks around both $0$ and $\pm\pi$ builds
up in the characteristic time $\gamma^{-1}$ needed to reach a
steady state \cite{Filipowicz:Theorymicromaser}. In the Josephson
regime, the relative phase locks around $0$ ($\pm\pi$), for
repulsive (attractive) two-body interactions, see
Fig.~\ref{figII5}(b). In contrast to these two regimes, in the
Fock regime the relative phase distribution becomes almost random
for all times, and the localized modes in the two wells evolve
independently of each other.

The three regimes of phase distributions correspond to different
orders of magnitude of the ratio $u_b/t_J$.  The crossover of the
non-equilibrium steady state from a phase-coherent regime to the
random-phase situation is reminiscent of the superfluid-Mott
insulator phase transition for the ground state of an optical
lattice \cite{Fisher:bosonlocalizationSFMI,Jaksch:BECinLattice}.
Since we consider just two sites, however, there is no sharp
transition between these regimes.

\section{\label{passagetimestatistics} Passage time statistics of molecule formation}

We now turn to a second-example that illustrates the understanding
of the dynamics of quantum-degenerate atomic and molecular systems
that can be gained from quantum optics analogies. Here, we
consider the first stages of coherent molecular formation via
photo-association. Since in such experiments the molecular field is
typically in a vacuum initially, it is to be intuitively expected
that the initial stages of molecule formation will be strongly
governed by quantum noise, hence subject to large fluctuations.
One important way to characterize these fluctuations is in terms
of the so-called passage time, which is the time it takes to
produce, or dissociate, a predetermined number of molecules.
Quantum noise results in fluctuations in that time, whose
probability distribution can therefore be used to probe the
fluctuations in the formation dynamics.

Because of the analogy between pairs of fermionic atoms and
two-level systems that we already exploited in the discussion of
the molecular micromaser, one can expect that the problem at hand
is somewhat analogous to spontaneous radiation from a sample of
two-level atoms, the well-know problem of superradiance. In this
section we show that this is indeed the case, and use this analogy
to study the passage time statistics of molecular formation from
fermionic atoms.

\subsection{Model}

We consider again a quantum-degenerate gas of fermionic atoms of
mass $m_f$ and spin $\sigma=\uparrow, \downarrow$, coupled
coherently to bosonic molecules of mass $m_b = 2m_f$ and zero
momentum via photo-association. Neglecting collisions between
fermions and assuming that for short enough times the molecules
occupy a single-mode of the bosonic field, this system can be
described by the boson-fermion model Hamiltonian

\begin{equation}
\label{Hamil1} H = \sum_{k}
\frac{1}{2}\hbar\omega_k\left(\hat{c}^\dagger_{k\uparrow}\hat{c}_{k\uparrow}
+
\hat{c}^\dagger_{-k\downarrow}\hat{c}_{-k\downarrow}\right)+\hbar\omega_b\hat{b}^\dagger
\hat{b} + \hbar\chi\sum_k\left(\hat{b}^\dagger
\hat{c}_{k\uparrow}\hat{c}_{-k\downarrow} +
\hat{b}\hat{c}^\dagger_{-k\uparrow}\hat{c}^\dagger_{k\downarrow}\right),
\end{equation}
where $\hat{b}^\dagger, \hat{b}$ are molecular bosonic creation
and annihilation operators and $\hat{c}_{k\sigma}^\dagger,
\hat{c}_{k\sigma}$ are fermionic creation and annihilation
operators describing atoms of momentum $\hbar k$ and spin
$\sigma$. The first and second terms in Eq.  (\ref{Hamil1})
describe the kinetic energy $\hbar\omega_k/2=\hbar^2k^2/(2m_f)$ of
the atoms and the detuning energy of the molecules respectively,
and the third term describes the photo-association of pairs of
atoms of opposite momentum into molecules.

Introducing the pseudo-spin operators \cite{Anderson:RPAinSuperconductivity} analogous to Eq. \eqref{andersonmapping}
\begin{eqnarray}
    \hat{\sigma}^z_k &=&
\frac{1}{2}(\hat{c}^\dagger_{k\uparrow}\hat{c}_{k\uparrow} +
    \hat{c}^\dagger_{-k\downarrow}\hat{c}_{-k\downarrow}-1),\nonumber \\
    \hat{\sigma}^+_k &=&(\hat{\sigma}^-_k)^\dagger=
    \hat{c}^\dagger_{-k\downarrow}\hat{c}^\dagger_{k\uparrow},
\label{A-mapping}
\end{eqnarray}
the Hamiltonian (\ref{Hamil1}) becomes, within an unimportant constant
\cite{Barankov,Meiser:singlemode_molecules},
\begin{equation}
\label{Hamil2} H = \sum_k \hbar\omega_k\hat{\sigma}^z_k
+\hbar\omega_b\hat{b}^\dagger \hat{b} +
\hbar\chi\sum_k\left(\hat{b}^\dagger\hat{\sigma}^-_k +
\hat{b}\hat{\sigma}^+_k\right).
\end{equation}
This Hamiltonian is known in quantum optics as the inhomogeneously
broadened (or non-degenerate) Tavis-Cummings
model~\cite{Tavis:Tavis_Cummings}. It describes the coupling of an
ensemble of two-level atoms to a single-mode electromagnetic
field. Hence the mapping (\ref{A-mapping}) establishes the formal
analogy between the problem at hand and Dicke superradiance, with
the caveat that we are dealing with a single bosonic
mode~\cite{Barankov,Meiser:singlemode_molecules,Javanainen,andreev04,Vardi04:dynamics,Taka05:BFC}.
Instead of real two-level atoms, pairs of fermionic atoms are now
described as effective two-level systems whose ground state
corresponds to the absence of a pair, $|g_k\rangle
=|0_{k\uparrow},0_{-k\downarrow}\rangle$ and the excited state to
a pair of atoms of opposite momenta, $|e_k\rangle =
|1_{k\uparrow},1_{-k\downarrow}\rangle$, in close analogy to the
treatment of the atoms in the previous section.

The initial condition consists of the molecular field in the
vacuum state and a filled Fermi sea of atoms
    \begin{equation}
    |F\rangle=\prod_{k\leq |k_F|} \hat{\sigma}^+_k |0\rangle,
    \end{equation}
where $k_F$ is the Fermi momentum. As such, the problem at hand is
in direct analogy to the traditional superradiance problem where
one starts from an ensemble of excited two-state atoms, as
expected from our previous comments. Later on we will also
consider an initial state containing only molecules and no atoms.
This is an important extension of the traditional Dicke
superradiance system, where the two-level atoms are coupled to all
modes of the photon vacuum a situation, thereby precluding the
possibility of an initial state containing a single,
macroscopically occupied photon mode unless the system us prepared
in a high-$Q$ cavity.

We assume from now on that the inhomogeneous broadening due to the
spread in atomic kinetic energies can be ignored. This so-called
degenerate approximation is justified provided that the kinetic
energies are small compared to the atom-molecule coupling energy,
$\beta=\epsilon_F/(\hbar \chi)\ll1$, where $\epsilon_F$ is the
Fermi energy.  It is the analog of the homogeneous broadening
limit of quantum optics, and of the Raman-Nath approximation in
atomic diffraction.  A comparison with typical experimental
parameters \cite{Heinzen:superchemistry} shows that the degenerate
approximation is justified if the number of atoms does not exceed
$\sim 10^2 - 10^3$ \cite{Hermann:passagetimestatistics}.

Limiting thus our considerations to small atomic samples, we
approximate all $\omega_k$'s by $\omega_F$ and introduce the
collective pseudo-spin operators
    \begin{eqnarray}
    \hat{S}_z = \sum_k \hat{\sigma}^z_k, \nonumber\\
    \hat{S}^\pm = \sum_k \hat{\sigma}^\pm_k,
    \end{eqnarray}
obtaining the standard Tavis-Cummings
Hamiltonian~\cite{Tavis:Tavis_Cummings,Taka05:BFC}
\begin{equation}
\label{Hamil3} H = \hbar\omega_F \hat{S}_z +\hbar\omega_b
\hat{b}^{\dagger}\hat{b} +\hbar\chi(\hat{b}\hat{S}^{+} +
\hat{b}^{\dagger}\hat{S}^-).
\end{equation}
This Hamiltonian conserves the total spin operator $\hat{\bm
S}^2$. The total number of atoms is twice the total spin and hence
is also a conserved quantity. $\hat{S}_z$ measures the difference
in the numbers of atom pairs and molecules.

Eq. (\ref{Hamil3}) can be diagonalized numerically with reasonable
computation times even for relatively large numbers of atoms. One
can, however, gain significant intuitive insight in the underlying
dynamics by finding operator equations of motion and then treating
the short-time molecular population semiclassically, $\langle{\hat
n}_b\rangle \rightarrow n_b$.  To this end we introduce the
``joint coherence'' operators
    \begin{eqnarray}
    \hat{T}_x &=& (\hat{b}\hat{S}^+ + \hat{b}^\dagger \hat{S}^-)/2, \nonumber \\
    \hat{T}_y &=& (\hat{b}\hat{S}^+ - \hat{b}^\dagger \hat{S}^-)/2i,
    \end{eqnarray}
and find the Heisenberg equations of motion
\begin{eqnarray}
\label{heisen1}
    \dot{\hat{n}}_b &=&-2\chi \hat{T}_y, \\
\label{heisen2}    \dot{\hat{T}}_x &=& \delta \hat{T}_y \\
\label{heisen3}    \dot{\hat{T}}_y &=& -\delta \hat{T}_x-\chi\left (2
\hat{S}_z\hat{n}_b +
\hat{S}^+\hat{S}^-\right ),
\end{eqnarray}
where $\delta = \omega_b-\omega_F$, so that $2\chi \hat{T}_x + \delta
\hat{n}_b$ is a constant of motion.

\begin{figure}
\begin{center}
\includegraphics[width=8cm,height=4.5cm]{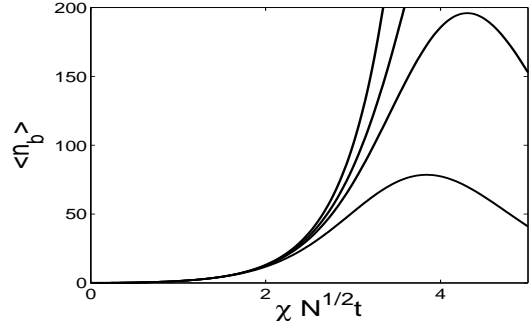}
\end{center}
\caption{Short-time dynamics of $\langle {\hat n}_b\rangle$. From
left to right, the curves give the linearized solution
(\ref{nbt2}) and the full quantum results for $N=500$, $N=250$,
and $N=100$, respectively. Figure taken from Ref. \cite{Hermann:passagetimestatistics}} \label{figIII1}
\end{figure}

In the following, we confine our discussion to the case of $\delta
= 0$ for simplicity. We thus neglect the contribution of $\hat{T}_x$ in
Eq. (\ref{heisen3}). In order to better understand the short time
dynamics we reexpress $\hat{S}^{+}\hat{S}^{-}$ as
\begin{equation}
\label{spm}
\hat{S}^+\hat{S}^- = -\hat{n}_b^2 + (2S-1)\hat{n}_b + N.
\end{equation}
This shows that the operator $\hat{S}^+\hat{S}^-$ is non-vanishing
when the molecular field is in a vacuum and hence can be
interpreted as a noise operator. Indeed
Eqs.~(\ref{heisen1})-(\ref{heisen3}) show that the buildup of the
molecular field is triggered only by noise if $\hat{n}_b=0$
initially. By keeping only the lowest-order terms in $\hat{n}_b$
we can eliminate $\hat{T}_y$ to obtain the differential equation
\begin{equation}
\ddot{\hat{n}}_b \approx 2N\chi^2\left
(2\hat{n}_b+1\right)
\end{equation}
which, for our initial state, may be solved to yield
\begin{equation}
\label{nbt2} \langle {\hat n}_b(t)\rangle \approx \sinh^2{(\chi\sqrt{N}t)}.
\end{equation}
Fig.~\ref{figIII1} compares the average molecule number $\langle
{\hat n}_b\rangle$ obtained this way, with the full quantum
solution obtained by direct diagonalization of the
Hamiltonian~(\ref{Hamil3}) for various values of $N$. The
semiclassical approach agrees within $5\%$ of the full quantum
solution until about $20\%$ of the population of atom pairs has
been converted into molecules in all cases.
\begin{figure}
\begin{center}
\includegraphics[width=8cm,height=4.5cm]{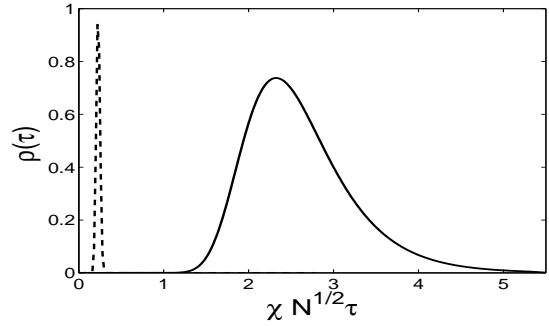}
\end{center}
\caption{Passage time distribution for converting 5$\%$ of the
initial population consisting of only atoms (molecules) into
molecules (atoms) for $N = 500$. For initially all atoms: solid
line, for initially all molecules: dashed line.} \label{figIII4}
\end{figure}

Next we turn to the passage time statistics.  In Fig~\ref{figIII4}
we show (solid line) the distribution of times required to produce
a normalized molecule number $n_b^{\rm ref}/N = 0.05$ from a
sample initially containing $N=500$ pairs of atomic fermions, as
found by direct diagonalization of the Hamiltonian (\ref{Hamil3}).
This distribution differs sharply from its counterpart for the
reverse process of photodissociation from a molecular condensate
into fermionic atom pairs, which is plotted as the dashed line in
Fig.~\ref{figIII4}. In contrast to photo-association, this latter
process suffers significantly reduced fluctuations.

\begin{figure}
\begin{center}
\includegraphics[width=8cm,height=4.5cm]{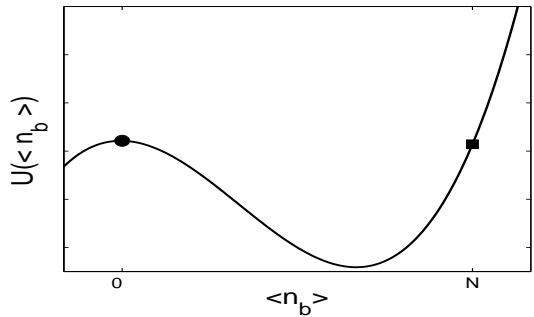}
\end{center}
\caption{Effective potential for a system with $N\gg1$. The
circle (square) corresponds to an initial state with all fermionic
atoms (molecules). The part of the potential for $n_b<0$ is unphysical.
Figure taken from Ref. \cite{Hermann:passagetimestatistics}}
\label{figIII5}
\end{figure}

To understand the physical mechanism leading to this reduction in
fluctuations we again turn to our short time semi-classical model.
Within this approximation, the Heisenberg equations of motion
(\ref{heisen1})-(\ref{heisen2}) can be recast in the form of a
Newtonian equation~\cite{Taka05:BFC}
\begin{equation}
\label{newton} \frac{d^2 n_b}{dt^2} = -\frac{d U(n_b)}{dn_b},
\end{equation}
where the cubic effective potential $U(n_b)$ is plotted in
Fig.~\ref{figIII5}. (Note we have now kept all orders in $n_b$.)
In case the system is initially composed solely of fermionic
atoms, $n_b(0)=0$, the initial state is dynamically unstable, with
fluctuations having a large impact on the build-up of $n_b$. In
contrast, when it consists initially solely of molecules, $n_b=N$,
the initial state is far from the point of unstable equilibrium,
and $n_b$ simply ``rolls down'' the potential in a manner largely
insensitive to quantum fluctuations. This is a consequence of the
fact that the bosonic initial state provides a mean field that is
more amenable to a classical description. Hence, while the early
stages of molecular dimer formation from fermionic atoms are
characterized by large fluctuations in formation times that
reflect the quantum fluctuations in the initial atomic state, the
reverse process of dissociation of a condensate of molecular
dimers is largely deterministic. The diminished fluctuations in
this reversed process is peculiar to the atom-molecule system and
not normally considered in the quantum optics analog of Dicke
superradiance.

\section{\label{countingstatistics} Counting statistics of molecular fields}

An important quantum mechanical characteristic of a quantum field
is its counting (or number) statistics. In this section we show
how the similarity of the coherent molecule formation with quantum
optical sum-frequency generation can be used to determine the
counting statistics of the molecular field. In particular we show
how the counting statistics depends on the statistics of the atoms
from which the molecules are formed. Besides being interesting in
its own right, such an analysis is crucial for an understanding of
several recent experiments that used a ''projection'' onto
molecules to detect BCS superfluidity in fermionic systems
\cite{Regal:BEC_BCS_crossover,Zwierlein:BEC_BCScrossover}. Our
work shows that the statistical properties of the resulting
molecular field indeed reflect properties of the initial atomic
state and are a sensitive probe for superfluidity.

As before, we restrict our discussion to a simple model in which
all the molecules are generated in a single mode.  We use time
dependent perturbation theory to calculate the number of molecules
formed after some time $t$, $n(t)$, as well as the equal-time
second-order correlation function $g^{(2)}(t,t)$. We also
integrate the Schr\"odinger equation numerically for small numbers
of atoms, which allows us to calculate the complete counting
statistics $P_n$.

\subsection{BEC}

Consider first a cloud of weakly interacting bosons well below the
condensation temperature $T_c$. It is a good approximation to
assume that all atoms are in the condensate, described by the
condensate wave function $\psi_0(x)$. The coupled system of atoms
and molecules is described by the effective two-mode Hamiltonian
\cite{Javanainen:Twomodemodel,Anglin:twomode}
\begin{equation}
\hat{H}_{\text{BEC}}=\hbar \delta \hat{b}^\dagger \hat{b} +
\hbar \chi \left(\hat{b}^\dagger \hat{c}^2 + \hat{b} \hat{c}^{\dagger
2}\right). \label{twomode_hamiltonian}
\end{equation}
where $\hat{b}$, $\hat{b}^\dagger$ and $\hat{c}$,
$\hat{c}^\dagger$ are the bosonic annihilation and creation
operators for the molecules and for the atoms in the condensate,
respectively, $\delta$ is the detuning between the molecular and
atomic level, and $\hbar\chi$ is the effective coupling constant.

\begin{figure}
\includegraphics{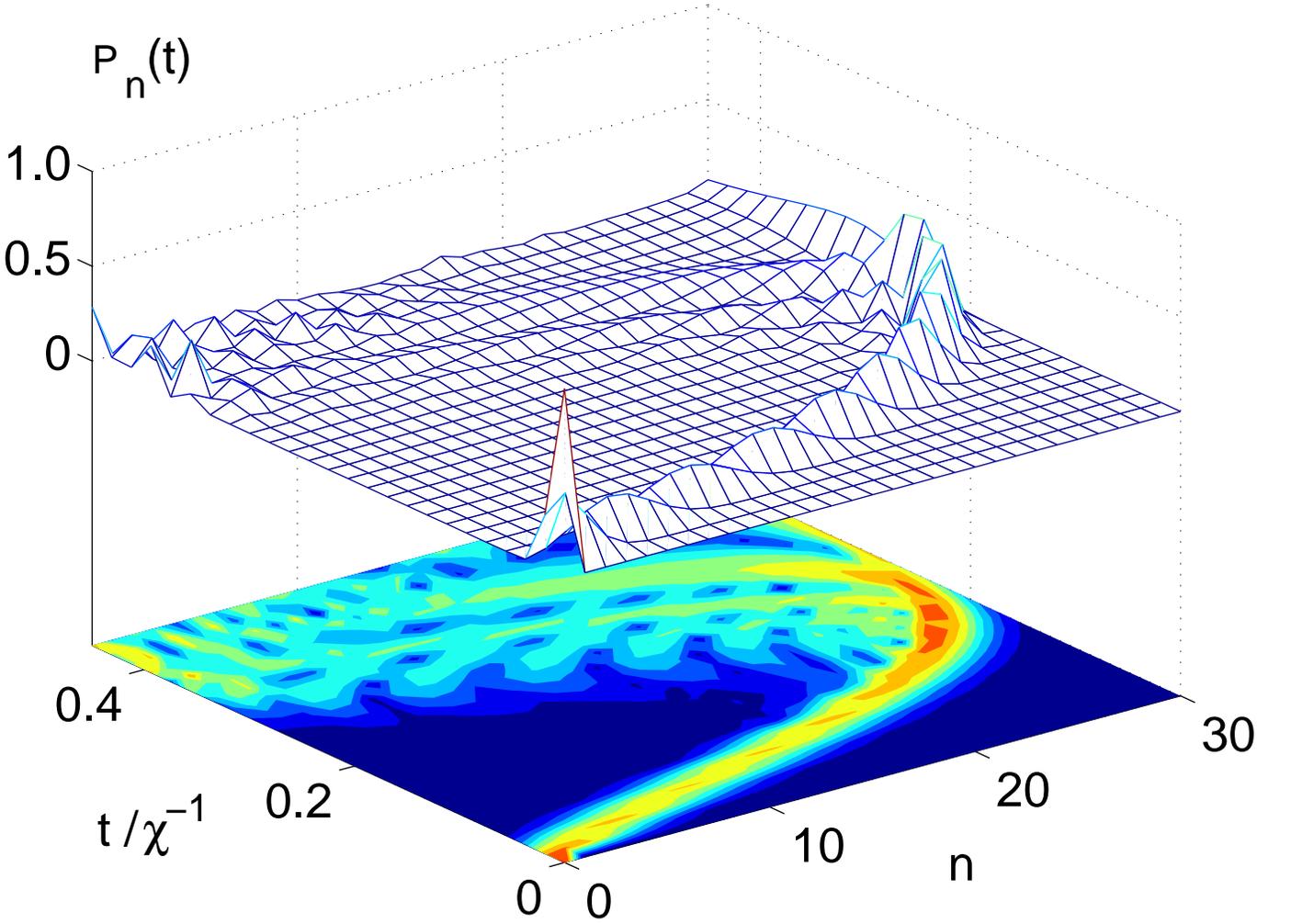}
\caption{Number statistics of molecules formed from a BEC with $N_{\rm max}=30$
and $\delta=0$.}
\label{PnBEC}
\end{figure}

Typical experiments start out with all atoms in the condensate and
no molecules, corresponding to the initial state,
\begin{equation}
\ket{\psi(t=0)}=\frac{\hat{c}^{\dagger{N_a}}}{\sqrt{N_a!}}\ket{0}.
\label{BEC_initial}
\end{equation}
where $N_a=2N_{\rm max}$ is the number of atoms, $N_{\rm max}$ is
the maximum possible number of molecules and $\ket{0}$ is the
vacuum of both molecules and atoms. We can numerically solve the
Schr\"odinger equation for this problem in a number basis and from
that solution we can determine the molecule statistics $P_n(t)$.
The results of such a simulation are illustrated in Fig.
\ref{PnBEC}, which shows $P_n(t)$ for 30 initial atom pairs and
$\delta=0$. Starting in the state with zero molecules, a
wave-packet-like structure forms and propagates in the direction
of increasing $n$. Near $N_{\rm max}$ the molecules begin to
dissociate back into atom pairs.

We can gain some analytical insight into the short-time dynamics
of molecule formation by using first-order perturbation theory
\cite{Kozierowski:FluctuationsSHG,Mandel:squeezingSHG}. We find
for the mean molecule number
\begin{equation}
n(t)=(\chi t)^2 2N_{\rm max}(2N_{\rm max}-1)+\Order{(\chi t)^2}
\label{nBEC}
\end{equation}
and for the second factorial moment
\begin{equation}
g^{(2)}(t_1,t_2) =1-\frac{2}{N_{\rm max}}+\Order{N_{\rm max}^{-2}}.
\end{equation}
For $N_{\rm max}$ large enough we have $g^{(2)}(t_1,t_2)
\rightarrow 1$, the value characteristic of a Glauber coherent
field. From $g^{(2)}$ and $n(t)$ we also find the relative width
of the molecule number distribution as
\begin{equation}
\frac{\sqrt{\langle(\hat{n}-n)^2\rangle}}{n}= \sqrt{g^{(2)} +
n^{-1}-1}. \label{deltanBEC}
\end{equation}
It approaches $n^{-1/2}$ in the limit of large $N_{\rm max}$,
typical of a Poisson distribution. This confirms that for short
enough times, the molecular field is coherent in the sense of
quantum optics.

\subsection{Normal Fermi gas}

We now turn to the case of photo-association from two different
species of non-interacting ultra-cold fermions. The two species
are again denoted by spin up and down. At $T=0$, the atoms fill a
Fermi sea up to an energy $\mu$. Weak repulsive interactions give
rise only to minor quantitative modifications that we ignore. We
refer to this system of non-interacting Fermions as a normal Fermi
gas (NFG) \cite{LandauIX}.

As before we assume that atom pairs are coupled only to a single
mode of the molecular field, which we assume to have zero momentum
for simplicity. Then, using the mapping to pseudo spins Eq.
\eqref{A-mapping} we find that the system is again described by
the inhomogeneously broadened Tavis-Cummings Hamiltonian Eq.
\eqref{Hamil2}. However, in contrast to the previous case, we do
not assume that the fermionic energies are approximately
degenerate, in order to be able compare the results to the BCS
case, where the kinetic energies are essential.

\begin{figure}
\includegraphics{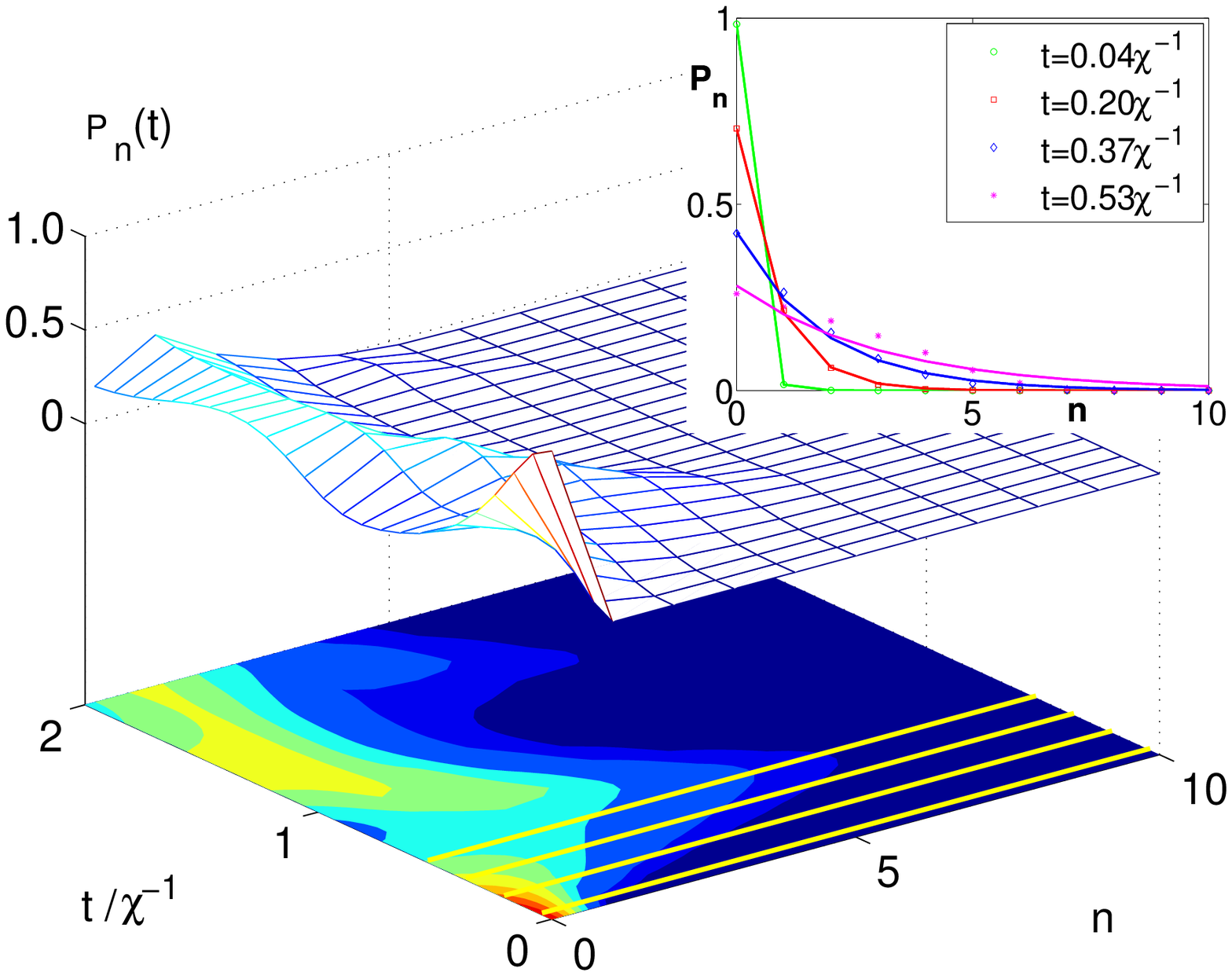}
\caption{Number statistics of molecules formed from a normal Fermi
gas. This simulation is for $N_a=20$ atoms, the detuning is
$\delta=0$, the Fermi energy is $\mu=0.1\hbar \chi$ and the
momentum of the $i$-th pair is $|k_i|=(i-1) 2 k_F/(N_a/2-1)$. The
inset shows fits of the number statistics to thermal distributions
for various times as marked by the thick lines in the main figure.}
\label{PnNFG}
\end{figure}

Figure \ref{PnNFG} shows the molecule statistics obtained this
way. The result is clearly both qualitatively and quantitatively
very different from the case of molecule formation from an atomic
BEC. From the Tavis-Cummings model analogy we expect that for
short times the statistics of the molecular field should be
chaotic, or ``thermal'', much like those of a single-mode chaotic
light field. This is because each individual atom pair ''emits'' a
molecule independently and without any phase relation with other
pairs. That this is the case is illustrated in the inset of Fig.
\ref{PnNFG}, which fits the molecule statistics at selected short
times with chaotic distributions of the form
\begin{equation}
P_{n,\text{thermal}}=\frac{e^{-n/\langle n\rangle}}{\sum_n
e^{-n/\langle n \rangle}} .\label{thermaldistribution_eqn}
\end{equation}
The increasing `pseudo-temperature' $\langle n \rangle$
corresponds to the growing average number of molecules as a
function of time.

As before we determine the short-time properties of the molecular
field in first-order perturbation theory.  We find for the mean
number of molecules
\begin{equation}
n(t)=(\chi t)^2 2N_{a}.
\end{equation}
It is proportional to $N_a$, in contrast to the BEC result, where
$n$ was proportional to $N_a^2$, see Eq.~(\ref{nBEC}). This is
another manifestation of the independence of all the atom pairs
from each other: While in the BEC case the molecule production is
a collective effect with contributions from all possible atom
pairs adding constructively, there is no such collective
enhancement in the case of Fermions. Each atom can pair up with
only one other atom to form a molecule. For the second factorial
moment we find
\begin{equation}
g^{(2)}(t_1,t_2)=2\left(1-\frac{1}{2N_a}\right)
\label{g2NFG}
\end{equation}
which is close to two, typical of a chaotic or thermal field.

\subsection{Fermi gas with superfluid component}

Unlike repulsive interactions, attractive interactions between
fermions have a profound impact on molecule formation. It is known
that such interactions give rise to a Cooper instability that
leads to pairing and drastically changes the qualitative
properties of the atomic system. The BCS reduced Hamiltonian is
essentially the inhomogeneously broadened Tavis-Cummings
Hamiltonian \eqref{Hamil2} with an additional term accounting for
the attractive interactions between atoms \cite{Kittel},
\begin{equation}
H = \sum_k \hbar\omega_k\hat{\sigma}^z_k
+\hbar\omega_b\hat{b}^\dagger \hat{b} +
\hbar\chi\sum_k\left(\hat{b}^\dagger\hat{\sigma}^-_k +
\hat{b}\hat{\sigma}^+_k\right)-V\sum_{k,k^\prime}
\hat{\sigma}_k^+ \hat{\sigma}_{k^\prime}^- .
\end{equation}
The approximate mean-field ground state $\ket{BCS}$ is found by
minimizing $\langle \hat{H}_\text{BCS}-\mu \hat{N}\rangle$ in the
standard way. The dynamics is then obtained by numerically
integrating the Schr\"odinger equation with $\ket{\text{BCS}}$ as
the initial atomic state and the molecular field in the vacuum
state.

\begin{figure}
\includegraphics{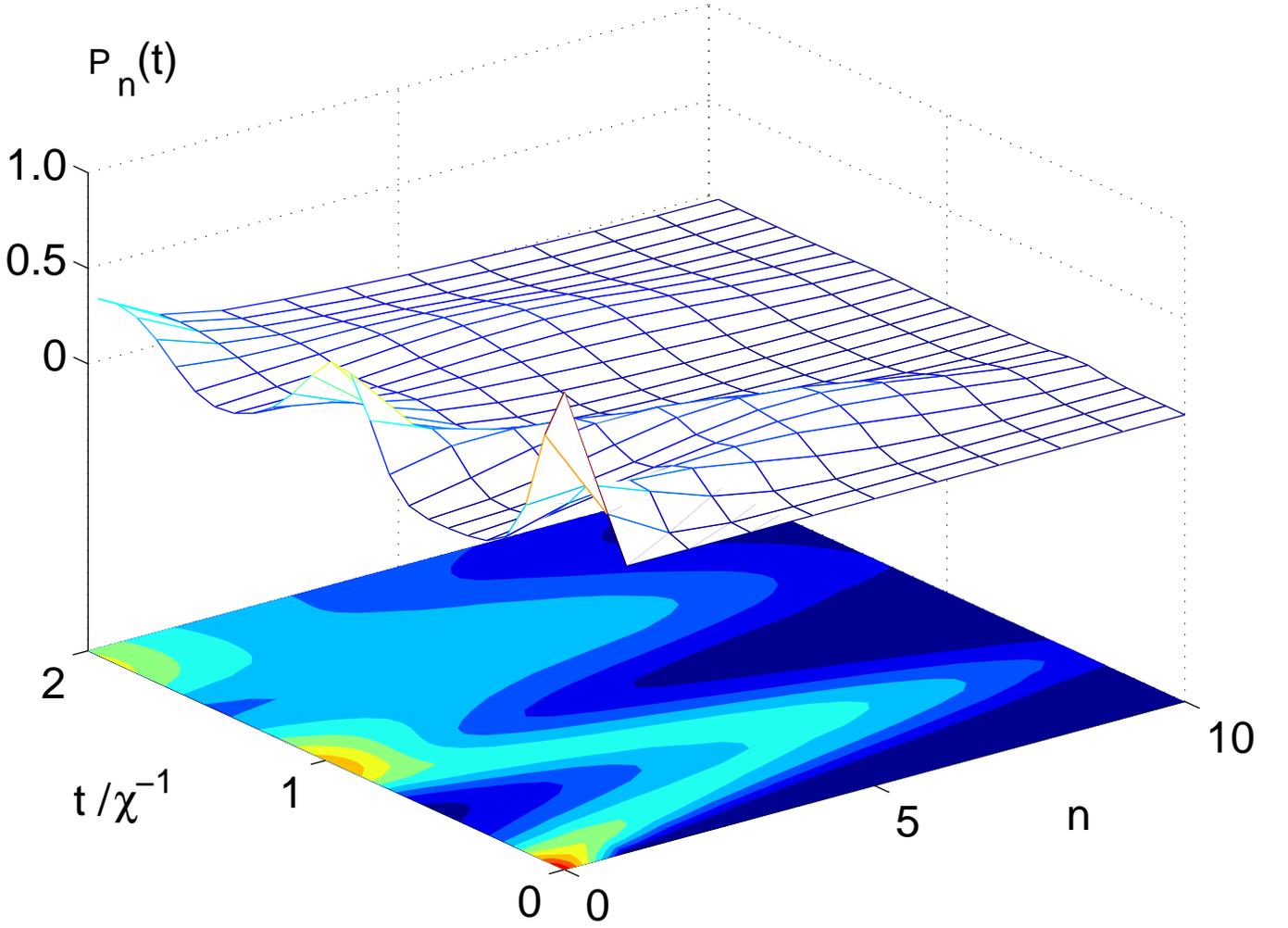}
\caption{Number statistics of molecules formed from a Fermi gas
with pairing correlations. For this simulation the detuning is
$\delta=0$, the Fermi energy is $\mu=0.1g$ and the background
scattering strength is $V=0.03\chi$ resulting in $N_a\approx 9.4$
atoms and a gap of $\Delta\approx0.15\chi$. The momenta of
the atom pairs are distributed as before in the normal Fermi gas
case.} \label{PnBCS}
\end{figure}

Figure \ref{PnBCS} shows the resulting molecule statistics for
$V=0.03\hbar \chi$, which corresponds to a gap of
$\Delta=0.15\hbar \chi$ for the system at hand. Clearly, the
molecule production is much more efficient than it was in the case
of a normal Fermi gas. The molecules are produced at a higher rate
and the maximum number of molecules is larger. The evolution of
the number statistics is reminiscent of the BEC case. This also
shows that the qualitative differences seen between the normal
Fermi gas and a BEC in the previous section cannot be attributed
to inhomogeneous broadening and the resulting dephasing alone but
are instead a result of the different coherence properties of the
atoms.

\begin{figure}
\includegraphics{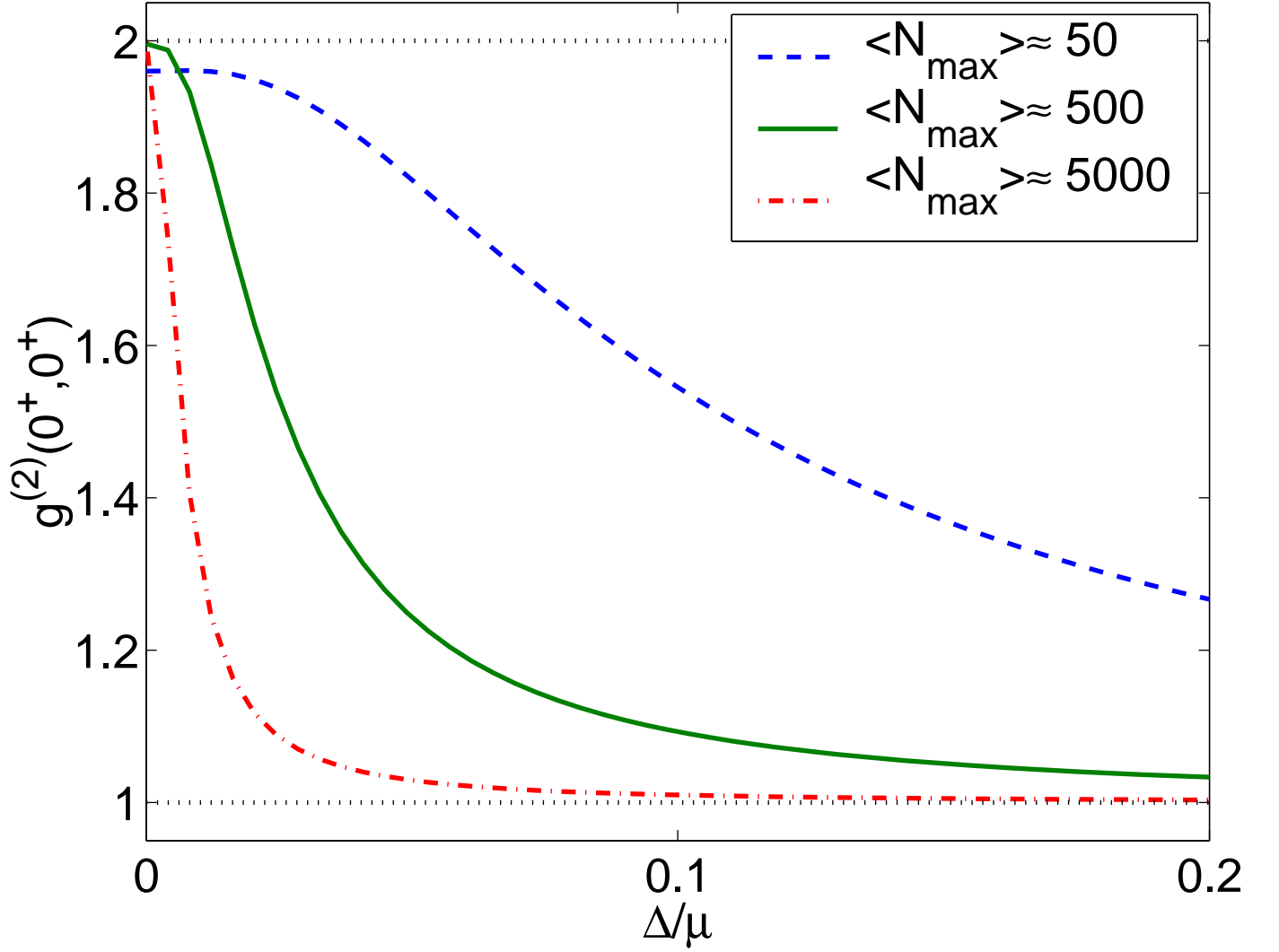}
\caption{$g^{(2)}(0^+,0^+)$ as a function of the gap parameter
$\Delta$. (Figure taken from Ref. \cite{Meiser:singlemode_molecules})} \label{g2ofgap}
\end{figure}

The short-time dynamics is again obtained in first-order
perturbation theory, which gives now
\begin{equation}
n(t)\approx(\chi t)^2\left[\left(\frac{\Delta}{V}\right)^2 + N_a\right].
\end{equation}
In addition to the term proportional to $N_a$ representing the
incoherent contribution from the individual atom pairs that was
already present in the normal Fermi gas, there is now an
additional contribution proportional to $(\Delta/V)^2$. Since
$(\Delta/V)$ can be interpreted as the number of Cooper pairs in
the quantum-degenerate Fermi gas, this term can be
understood as resulting from the {\em coherent} conversion of
Cooper pairs into molecules in a collective fashion similar to the
BEC case. The coherent contribution results naturally from the
nonlinear coupling of the atomic field to the molecular field.
This nonlinear coupling links higher-order correlations of the
molecular field to lower-order correlations of the atomic field.
For the parameters of Fig. \ref{PnBCS} $\Delta/V\approx 6.5$ so
that the coherent contribution from the Cooper pairs clearly
dominates over the incoherent contribution from the unpaired
fermions. Note that no signature of that term can be found in the
momentum distribution of the atoms themselves. Their momentum
distribution is very similar to that of a normal Fermi gas. The
short-time value of $g^{(2)}(t_1,t_2)$, shown in Fig.
\ref{g2ofgap}, decreases from the value of Eq. \eqref{g2NFG} for a normal
Fermi gas at $\Delta=0$ down to one as $\Delta$ increases,
underlining the transition from incoherent to coherent molecule
production.

\section{\label{diagnosingwithmolecules} Molecules as probes of spatial correlations}

The single-mode description of the molecular field of the previous
section results in the loss of all information about the spatial
structure of the atomic state. In this final section we adopt a
complementary view and study the coupled atom-molecule system
including all modes of the molecular and atomic field so as to
resolve their spatial structure. This problem is too complex to
admit an exact solution, hence we rely entirely on perturbation
theory.

One of the motivations for such studies is the on-going
experimental efforts to study the so-called BEC-BCS crossover. A
difficulty of these studies has been that they necessitate the
measurement of higher-order correlations of the atomic system.
While the momentum distribution of a gas of bosons provides a
clear signature of the presence of a Bose-Einstein condensate, the
Cooper pairing between fermionic atoms in a BCS state hardly
changes the momentum distribution or spatial profile as compared
to a normal Fermi gas. This poses a significant experimental
challenge, since the primary techniques for probing the state of
an ultra-cold gas are either optical absorption or phase contrast
imaging, which directly measure the spatial density or momentum
distribution following ballistic expansion of the gas. In the
strongly interacting regime very close to the Feshbach resonance,
evidence for fermionic superfluidity was obtained by projecting
the atom pairs onto a molecular state by a rapid sweep through the
resonance
\cite{Regal:BEC_BCScrossover,Zwierlein:BEC_BCScrossover}. More
direct evidence of the gap in the excitation spectra due to
pairing was obtained by rf spectroscopy \cite{Chin:} and by
measurements of the collective excitation frequencies
\cite{Kinast:,Bartenstein:}. Finally, the superfluidity of
ultra-cold fermions in the strongly interacting regime has recently
been impressively demonstrated via the generation of atomic
vortices \cite{Zwierlein:BCSvortices}.

Still, the detection of fermionic superfluidity in the weakly
interacting BCS regime remains a challenge. The direct detection
of Cooper pairing requires the measurement of second-order or
higher atomic correlation functions. Several researchers have
proposed and implemented schemes that allow one to measure higher
order correlations
\cite{Burt:BEC_decay,Hellweg:Spatial_correlations,Cacciapuoti:Second_correlation_function,Altman:noise_correlations,Regal:BEC_BCScrossover,Radka:Diagnosing_correlations}
but those methods are still very difficult to realize
experimentally.
While the measurement of higher-order correlations is challenging
already for bosons, the theory of these correlations has been
established a long time ago by Glauber for photons
\cite{Glauber:Optical_Coherence1,Glauber:Optical_Coherence2,Naraschewski:Spatial_coherence}.
For fermions however, despite some efforts
\cite{Cahill:Density_Operator_Fermions} a satisfactory coherence
\emph{theory} is still missing.

From the previous section we know that one can circumvent these
difficulties by making use of the nonlinear coupling of atoms to a
molecular field. The nonlinearity of the coupling links
first-order correlations of the molecules to second-order
correlations of the atoms. Furthermore the molecules are always
bosonic so that the well-known coherence theory for bosonic fields
can be used to characterize them.  Considering a simplified model
with only one molecular mode, it was found that the molecules
created that way can indeed be used as a diagnostic tool for
second-order correlations of the original atomic field.

We consider the limiting case of strong atom-molecule coupling as compared to
the relevant atomic energies. The molecule formation from a Bose-Einstein
condensate
(BEC) serves as a reference system. There we can rather easily
study the contributions to the molecular signal from the condensed
fraction as well as from thermal and quantum fluctuations above
the condensate. The cases of a normal Fermi gas and a BCS
superfluid Fermi system are then compared with it. We show that
the molecule formation from a normal Fermi gas and from the
unpaired fraction of atoms in a BCS state has very similar
properties to those of the molecule formation from the non-condensed
atoms in the BEC case. The state of the molecular field formed
from the pairing field in the BCS state on the other hand is
similar to that resulting from the condensed fraction in the BEC
case. The qualitative information gained by the analogies with the
BEC case help us gain a physical understanding of the molecule
formation in the BCS case where direct calculations are difficult
and not nearly as transparent.

\subsection{\label{model} Model}

We consider again the three cases where the atoms are bosonic and
initially form a BEC, or consist of two species of ultra-cold
fermions (labeled again by $\sigma=\uparrow,\downarrow$), with or
without superfluid component. In the following we describe
explicitly the situation for fermions, the bosonic case being
obtained from it by omitting the spin indices and by replacing the
Fermi field operators by bosonic field operators.

Since we are primarily interested in how much can be learned about
the second-order correlations of the initial atomic cloud from the
final molecular state, we keep the physics of the atoms themselves
as well as the coupling to the molecular field as simple as
possible. The coupled fermion-molecule system can be described by
the Hamiltonian
\cite{Holland:resonant_superfluidity,Chiofalo:Res_superfluidity,Timmermans:Feshbachresonances}
\begin{widetext}
\begin{eqnarray}
\hat{H}&=&\sum_{\vec{k},\sigma}
\frac{\omega_\vec{k}}{2}\hat{c}_{\vec{k}\sigma}^\dagger
\hat{c}_{\vec{k}\sigma}+
\sum_\vec{k}\omega_\vec{k}\hat{a}_{\vec{k}}^\dagger \hat{a}_{\vec{k}}+
V^{-1/2}\sum_{\vec{k}_1,\vec{k}_2,\sigma}
\tilde{U}_\text{tr}(\vec{k}_2-\vec{k}_1)
\hat{c}_{\vec{k}_2\sigma}^\dagger
\hat{c}_{\vec{k}_1\sigma}\nonumber \\
&+& \frac{U_0}{2V}\sum_{\vec{q},\vec{k}_1,\vec{k}_2}
\hat{c}_{\vec{k}_1+\vec{q}\uparrow}^\dagger
\hat{c}_{\vec{k}_2-\vec{q}\downarrow}^\dagger
\hat{c}_{\vec{k_2}\downarrow} \hat{c}_{\vec{k}_1\uparrow}+
\hbar g\left(\sum_{\vec{q},\vec{k}}\hat{a}_{\vec{q}}^\dagger
\hat{c}_{\vec{q}/2+\vec{k}\downarrow}\hat{c}_{\vec{q}/2-\vec{k}\uparrow}
+ \text{H.c.}\right) \label{full_hamiltonian}
\end{eqnarray}
\end{widetext}
The kinetic energies $\omega_\vec{k}$ are defined as before, $V$ is the
quantization volume, $\tilde{U}_\text{tr}(\vec{k})=V^{-1/2}\int_V
d^3x e^{-i\vec{k}\vec{x}} U_{\rm tr}(\vec{x})$ is the Fourier
transform of the trapping potential $U_{\rm tr}({\bf r})$ and $U_0
= 4\pi \hbar^2 a/m_f$ is the background scattering strength with
$a$ the background scattering length. The coupling constant $g$
between atoms and molecules is, up to dimensions, equal to $\chi$
of the previous sections.

We assume that the trapping potential and background scattering
are relevant only for the preparation of the initial state before
the coupling to the molecules is switched on at $t=0$ and can be
neglected in the calculation of the subsequent dynamics. This is
justified if $\hbar g\sqrt{N} \gg U_0n,\hbar \omega_i$ where $n$
is the atomic density, $N$ the number of atoms, and $\omega_i$ are
the oscillator frequencies of the atoms in the potential $U_{\rm
tr}({\bf r})$ that is assumed to be harmonic.  Experimentally, the
interaction between the atoms can effectively be switched off by
ramping the magnetic field to a position where the scattering
length is zero, so that this assumption is fulfilled.

Regarding the strength of the coupling constant $g$, two cases are
possible: $\hbar g\sqrt{N}$ can be much larger or much smaller
than the characteristic kinetic energies involved. For fermions
the terms broad and narrow resonance have been coined for the two
cases, respectively, and we will use these for bosons as well.
Both situations can be realized experimentally, and they give rise
to different effects. For strong coupling the conversion process
needs not satisfy energy conservation because of the energy time
uncertainty relation. For weak coupling energy conservation is
enforced. This energy selectivity can be useful in certain situations because it
allows one to resolve additional structures in the atomic state.
The analysis of this case is fairly technical, however. Therefore
we only consider the case  of strong coupling and refer the
interested reader to \cite{Meiser:diagnosingwithmolecules} for
details of the calculations and the case of weak coupling.

First-order time-dependent perturbation theory requires that the
state of the atoms does not change significantly and consequently,
only a small fraction of the atoms are converted into molecules.
It is reasonable to assume that this is true for short interaction
times or weak enough coupling. Apart from making the system
tractable by analytic methods there is also a deeper reason why
the coupling should be weak: Since we ultimately wish to get
information about the atomic state, it should not be modified too
much by the measurement itself, i.e. the coupling to the molecular
field. Our treatment therefore follows the same spirit as
Glauber's original theory of photon detection, where it is assumed
that the light-matter coupling is weak enough that the detector
photocurrent can be calculated using Fermi's Golden rule.

\subsection{\label{BEC} BEC}

We consider first the case where the initial atomic state is a BEC
in a spherically symmetric harmonic trap. We assume that the temperature is
well below the BEC transition temperature and that the interactions between the
atoms are not too strong. Then the atomic system is described by the
field operator
\begin{equation}
\hat{\psi}(\vec{x})=\psi_0(\vec{x})\hat{c}+\delta\hat{\psi}(\vec{x}),
\label{decomposition_psi}
\end{equation}
where $\psi_0(\vec{x})$ is the condensate wave function
and $\hat{c}$ is the annihilation operator for an atom in the
condensate. In accordance with the assumption of low
temperatures and weak interactions we do not distinguish
between the total number of atoms and the number of atoms in the
condensate. The fluctuations $\delta \hat{\psi}(\vec{x})$ are small
and those with wavelengths much less than $R_{TF}$ will be treated
in the local density approximation while those with wavelengths
comparable to $R_{TF}$ can be neglected
\cite{Hutchinson:Finite_T_BEC,Reidl:Finite_T_BEC,Bergeman:BEC_Tc}.

We are interested in the momentum distribution of the molecules
\begin{equation}
n(\vec{p},t)=\langle \hat{b}_\vec{p}^\dagger(t) \hat{b}_\vec{p}(t)\rangle
\end{equation}
which for short times, $t$, can be calculated using perturbation
theory. In the broad resonance limit we ignore the kinetic
energies and find
\begin{widetext}
\begin{equation}
\label{g1bec}
n_{\rm BEC}(\vec{p},t)=
(gt)^2 N(N-1)V\left|{\tilde{\psi}_0^2}(\vec{p})\right|^2
 +(gt)^2 4N\int \frac{d^3x}{V}
 \left\langle \delta\hat{c}_\vec{p}^\dagger(\vec{x}) \delta\hat{c}_\vec{p}(\vec{x})\right\rangle,
\end{equation}
\end{widetext}
where the expectation value in the last term includes a thermal
average. From this expression we see that our approach is
justified if $(\sqrt{N}gt)^2\ll 1$ because for such times the
initial atomic state can be assumed to remain undepleted. The
first term in Eq. \eqref{g1bec} is the contribution from condensed
atoms and the second term comes from uncondensed atoms above the
condensate. The contribution from the condensate can be evaluated
in closed form in the Thomas Fermi approximation for a spherical
trap. The contribution from the thermal atoms can be calculated
using the local density approximation. The details of this
calculation can be found in Ref.
\cite{Meiser:diagnosingwithmolecules}.

\begin{figure}
\includegraphics{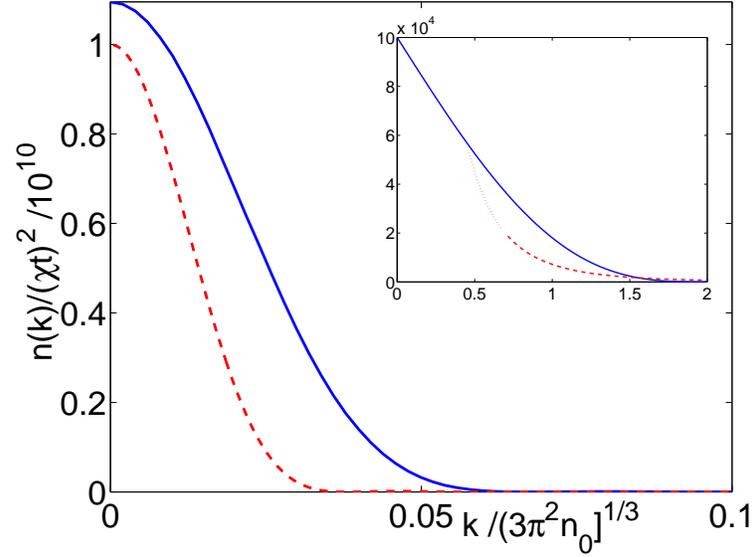}
\caption{Momentum distribution of molecules formed
from a BEC (red dashed line) with $a=0.1a_{\rm osc}$ and
$T=0.1T_c$ and a BCS type state with $k_F a=0.5$ and $a_{\rm
osc}=5k_F^{-1}(0)$ (blue solid line), both for $N=10^5$ atoms. The
BCS curve has been scaled up by a factor of $20$ for easier
comparison. The inset shows the noise contribution for BEC (red
dashed) and BCS (blue) case.  The latter is simply the momentum
distribution of molecules formed from a normal Fermi gas. The
local density approximation treatment of the noise contribution in
the BEC case is not valid for momenta smaller than $2\pi/\xi$
(indicated by the red dotted line in the inset). Note that the
coherent contribution is larger than the noise contribution by
five orders of magnitude in the BEC case and three orders of
magnitude in the BCS case. }\label{nofp_broad}
\end{figure}

The momentum distribution \eqref{g1bec} is illustrated in Fig.
\ref{nofp_broad}. The contribution from the condensate is a
collective effect, as indicated by its quadratic scaling with the
atom number. It clearly dominates over the incoherent contribution
from the fluctuations, which is proportional to the number of
atoms and only visible in the inset. The momentum width of the
contribution from the condensate is roughly $\hbar 2\pi/R_{\rm
TF}$ which is much narrower than the contribution from the
fluctuations, whose momentum distribution has a typical width of
$\hbar/\xi$, where $\xi=(8\pi a n)^{-1/2}$ is the healing length.
This is a case where coherence properties of the atoms can be read
off the momentum distribution of the molecules: The narrow
momentum distribution of the molecules is only possible if the
atoms were coherent over distances $\sim R_{TF}$. At this point
this is a fairly trivial observation and the same information
could have been gained by looking directly at the momentum
distribution of the atoms, which is after all how Bose-Einstein
condensation was detected already in the very first experiments
\cite{Cornell:BEC1995,Ketterle:BEC1995,Hulet:BEC1995}. Still we
mention it because it will be very interesting (indeed interesting
enough to motivate this whole work!) to contrast this situation to
the BCS case below.

Using the same approximation scheme we can calculate the second-order
correlation.  If we neglect fluctuations we find
\begin{equation}
g_{\rm BEC}^{(2)}(\vec{p}_1, t_1; \vec{p}_2, t_2)
=1-\frac{6}{N}+\Order{N^{-2}}.
\label{g2perturbative_bec}
\end{equation}
For $N\rightarrow \infty$ this is very close to 1, which is
characteristic of a coherent state. This
result implies that the number fluctuations of the molecules are
very nearly Poissonian. The fluctuations lead to a larger value of
$g^{(2)}$, making the molecular field partially coherent, but
their effect is only of order $\Order{N^{-1}}$.

\subsection{\label{NFG} Normal Fermi gas}

We treat the gas in the local density approximation where the atoms locally
fill a Fermi sea
\begin{equation}
\ket{NFG}=\prod_{|\vec{k}|<k_F(\vec{x})} \hat{c}_\vec{k}^\dagger
\ket{0}
\end{equation}
with local Fermi momentum $\hbar k_F(\vec{x})$ and $\ket{0}$ being the atomic
vacuum. It is related to the local density of the atoms in the usual way
\cite{LandauIX,Butts}.

The momentum distribution and second order correlation function are readily
found in perturbation theory. The momentum distribution is shown in the inset
in Fig. \ref{nofp_broad}. The total number of molecules scales only linear with
the number of atoms, meaning that, in contrast to the BEC case, the molecule
formation is non-collective. Each atom pair is converted into a molecule
independently of all the others and there is no collective enhancement.
Furthermore the momentum distribution of the atoms is much wider than in the
BEC case. It's width is of the order of $\hbar n_0^{1/3}$ indicating that the
atoms are correlated only over distances comparable to the inter atomic
distance.

Similarly, we find for the local value of $g^{(2)}$ at position $\vec{x}$
\begin{eqnarray}
g^{(2)}_{\rm loc}(\vec{p},\vec{x},t)&\equiv&
g^{(2)}_{\rm loc}(\vec{p},t;\vec{p},t,\vec{x})\nonumber\\
&=& 2\left(1-\frac{1}{N_{\rm eff}(\vec{p},\vec{x})}\right),
\label{g2fermigas}
\end{eqnarray}
where $N_{\rm eff}$ is the number of atoms that are allowed to form a molecule
on the basis of momentum conservation. For large $N_{\rm eff}$ $g^{(2)}$
approaches 2 which is characteristic of a thermal field. Indeed, using the
analogy with an ensemble of two level atoms coupled to every mode of the
molecular field provided by the Tavis-Cummings model, it is easy to see that
the entire counting statistics is thermal.

\subsection{\label{BCS} BCS state}

Let us finally consider a system of fermions with attractive
interactions, $U_0<0$, at temperatures well below the BCS critical
temperature. As is well known, for these temperatures the
attractive interactions give rise to correlations between pairs of
atoms in time reversed states known as Cooper pairs. We assume
that the spherically symmetric trapping potential is sufficiently
slowly varying that the gas can be treated in the local density
approximation. More quantitatively, the local density
approximation is valid if the size of the Cooper pairs, given by
the correlation length
\[
\lambda(r)= v_F(r)/\pi\Delta(r),
\]
is much smaller than the oscillator length for the trap. Here,
$v_F(r)$ is the velocity of atoms at the Fermi surface and
$\Delta(r)$ is the pairing field at distance $r$ from the origin,
which we take at the center of the trap. Loosely speaking, in the
local density approximation the ground state of the atoms is
determined by repeating the variational BCS-calculation of the
previous section in a small volume at every position $\vec{x}$. A
thorough discussion of this calculation can be found in Ref.
\cite{Houbiers:BCS_LDA}.

We find the momentum distribution of the molecules from the
BCS-type state by repeating the calculation done in the case of a
normal Fermi gas. For the BCS wave function, the relevant atomic
expectation values factorize into normal and anomalous
correlations. The normal terms are proportional to densities and
are already present in the case of a normal Fermi gas while the
anomalous contributions are proportional to the gap parameter. The
momentum distribution of the molecules becomes
\begin{equation}
n_{\rm BCS}(\vec{p},t)
\approx (gt)^2\left|\sum_k \Big\langle
\hat{c}_{\vec{p}/2+\vec{k},\downarrow}\hat{c}_{\vec{p}/2-\vec{k},\uparrow}
\Big\rangle\right|^2 + n_{\rm NFG}(\vec{p},t). \label{nofpBCS}
\end{equation}

The first term is easily shown to be proportional to the square of
the Fourier transform of the gap parameter. Since the gap
parameter is slowly varying over the size of the atomic cloud,
this contribution has a width of the order of $\hbar/R_{TF}$, in
complete analogy with the BEC case above. The total number of
atoms in the first contribution is proportional to the square of
the number of Cooper pairs, which is a macroscopic fraction of the
total atom number well below the BCS transition temperature. That
means that this contribution is a collective effect. The second
term is the wide and incoherent non-collective contribution
already present in the case of a normal Fermi gas. It is very
similar to the thermal noise in the BEC case as far as its
coherence properties are concerned.

For weak interactions such that the coherent contribution is small
compared to the incoherent contribution, the second order
correlations are close to those of a normal Fermi gas given by Eq.
\eqref{g2fermigas}, $g^{(2)}(\vec{p},\vec{x},t)\approx 2$.
However, in the strongly interacting regime, $k_F|a|\sim 1$, and
large $N$, the coherent contribution from the paired atoms
dominates over the incoherent contribution from unpaired atoms. In
this limit one finds that the second-order correlation is close to
that of the BEC, $g^{(2)}(\vec{p},\vec{x},t)\approx 1$. The
physical reason for this is that at the level of even order
correlations the pairing field behaves just like the mean field of
the condensate. This is clear from the factorization property of
the atomic correlation functions in terms of the normal component
of the density and the anomalous density contribution due to the
mean field. In this case, the leading order terms in $N$ are given
by the anomalous averages.

To summarize, molecules produced from an atomic BEC show a rather
narrow momentum distribution that is comparable to the zero-point
momentum width of the BEC from which they are formed. The molecule
production is a collective effect with contributions from all atom
pairs adding up constructively, as indicated by the quadratic
scaling of the number of molecules with the number of atoms.  Each
mode of the resulting molecular field is to a very good
approximation coherent (up to terms of order $\Order{1/N}$). The
effects of noise, both due to finite temperatures and to vacuum
fluctuations, are of relative order $\Order{1/N}$. They slightly
increase the $g^{(2)}$ and cause the molecular field in each
momentum state to be only partially coherent.

In contrast, the momentum distribution of molecules formed from a
normal Fermi gas is much broader with a typical width given by the
Fermi momentum of the initial atomic cloud, i.e. the atoms are
only correlated over an interatomic distance. The molecule
production is not collective as the number of molecules only
scales like the number of atoms rather than the square. In this
case, the second-order correlations of the molecules exhibit
super-Poissonian fluctuations, and the molecules are well
characterized by a thermal field.

The case where molecules are produced from paired atoms in a
BCS-like state shares many properties with the BEC case: The
molecule formation rate is collective, their momentum distribution
is very narrow, corresponding to a coherence length of order
$R_{TF}$, and the molecular field is essentially coherent. The
non-collective contribution from unpaired atoms has a momentum
distribution very similar to that of the thermal fluctuations in
the BEC case.

\section{\label{summary} Conclusion}

In this paper we have used three examples to illustrate the
profound impact of quantum optics paradigms, tools and techniques,
on the study of low-density, quantum-degenerate atomic and
molecular systems. There is little doubt that the remarkably fast
progress witnessed by that field results in no little part from
the experimental and theoretical methods developed in quantum
optics over the last decades. It is therefore fitting, on the
occasion of Herbert Walther's seventieth birthday, to reflect on
the profound impact of the field that he has helped invent, and
where he has been and remains so influential, on some of the most
exciting developments in AMO science.

\acknowledgments This work was supported in part by the US Office
of Naval Research, by the National Science Foundation, by the US
Army Research Office, and by the National Aeronautics and Space
Administration.

\bibliography{SPAM}

\end{document}